\documentclass[useAMS,usenatbib]{mn2e}
\usepackage{epsfig}
\usepackage{amssymb}
\usepackage{amsmath}
\usepackage{lscape}
\usepackage{longtable}
\usepackage[normal]{caption}
\usepackage{appendix}
\usepackage{bm}

\title[AGN are cooler than you think: the intrinsic far-IR emission
from QSOs]{AGN are cooler than you think: the intrinsic far-IR emission
from QSOs}

\author[M.~Symeonidis et al.] 
{\parbox{\textwidth}{\raggedright
M.~Symeonidis,$^{1,2}$\thanks{E-mail: \texttt{m.symeonidis@ucl.ac.uk}}
B. M. ~Giblin,$^{3,2}$
M. J. ~Page,$^{1}$ 
C. ~Pearson,$^{4}$
G. ~Bendo,$^{5}$
N. ~Seymour,$^{6}$
and S. J. ~Oliver $^{2}$
}\vspace{0.4cm}\\
\parbox{\textwidth}{\raggedright $^{1}$ Mullard Space Science
  Laboratory, University College London, Holmbury St. Mary, Dorking,
  Surrey RH5 6NT, UK\\
$^{2}$ Astronomy Centre, Dept. of Physics \& Astronomy, University of Sussex, Brighton BN1 9QH, UK\\
$^{3}$ University of Birmingham, Edgbaston, Birmingham, West Midlands B15 2TT\\
$^{4}$ RAL Space, Rutherford Appleton Laboratory, Chilton, Didcot, Oxfordshire OX11 0QX, UK\\
$^{5}$ UK ALMA Regional Centre Node, Jodrell Bank Centre for Astrophysics, Alan Turing Building, School of Physics and Astronomy, The University of Manchester, Oxford Road, Manchester, M13 9PL\\
$^{6}$ Department of Physics and Astronomy, Curtin University, Kent Street, Bentley, Perth, Western Australia, 6102.
}}

\begin{document}

\date{Accepted  Received; in original form}

\pagerange{\pageref{firstpage}--\pageref{lastpage}} \pubyear{2014}

\maketitle

\label{firstpage}

\begin{abstract}
We present an \textit{intrinsic} AGN SED extending from the optical to the submm, derived with a sample of unobscured, optically luminous ($\nu L_{\nu, 5100}$$>$ 10$^{43.5}$ erg/s) QSOs at $z<0.18$ from the Palomar Green survey. The intrinsic AGN SED was computed by removing the contribution from stars using the 11.3$\mu$m polycyclic aromatic hydrocarbon (PAH) feature in the QSOs' mid-IR spectra; the 1$\sigma$ uncertainty on the SED ranges between 12 and 45 per cent as a function of wavelength and is a combination of PAH flux measurement errors and the uncertainties related to the conversion between PAH luminosity and star-forming luminosity. Longwards of 20$\mu$m the shape of the intrinsic AGN SED is independent of the AGN power indicating that our template should be applicable to all systems hosting luminous AGN ($\nu L_{\nu, 5100}$ or $L_{\rm X(2-10keV)}$ $\gtrsim$10$^{43.5}$ erg/s). We note that for our sample of luminous QSOs, the average AGN emission is at least as high as, and mostly higher than, the total stellar powered emission at all wavelengths from the optical to the submm. This implies 
that in many galaxies hosting powerful AGN, there is no `safe' broadband photometric observation (at $\lambda<1000\mu$m) which can be used in calculating star-formation rates without subtracting the AGN contribution. Roughly, the AGN contribution may be ignored only if the intrinsic AGN luminosity at 5100$\rm \AA$ is at least a factor of 4 smaller than the total infrared luminosity ($L_{\rm IR}$; 8-1000$\mu$m) of the galaxy. Finally, we examine the implication of our work in statistical studies of star-formation in AGN host galaxies. 
\end{abstract}


\section{Introduction}
\label{sec:introduction}

The debate about a possible connection between galaxies and their supermassive black holes has been long-standing, originally fuelled by two global observations: (i) the correlation between black hole mass 
and bulge stellar mass or velocity dispersion in the nearby Universe (e.g. Magorrian et al. 1998\nocite{Magorrian98}; Ferrarese $\&$ Merritt 2000\nocite{FM00}) and (ii) the comparable temporal evolution of the 
star-formation rate density and AGN accretion rate density (e.g. Boyle $\&$ Terlevich 1998\nocite{BT98}). The idea that star-formation and AGN accretion are causally connected has become an essential 
component in theoretical models of galaxy evolution, as AGN feedback is required to regulate the formation and number of massive galaxies (e.g. Croton et al. 2006\nocite{Croton06}, Bower et al. 
2006\nocite{Bower06}). 

The connection between AGN and their hosts has often been examined by means of the location of galaxies in colour-magnitude space (e.g. Strateva et al. 2001\nocite{Strateva01}; Baldry et al. 
2004\nocite{Baldry04}). The sparsity of galaxies in transition between the `blue cloud' of star-forming galaxies and the `red sequence' of passive ellipticals has founded hypotheses in which energetic events 
associated with the central black hole are responsible for terminating star-formation on short timescales (e.g. Granato et al. 2004\nocite{Granato04}; Springel et al. 2005\nocite{SdMH05}). Studies have 
examined, albeit inconclusively, whether the colours of AGN hosts are an indication of the quenching of star formation; Georgakakis et al. (2008)\nocite{Georgakakis08a} and Schawinski et al. 
(2009)\nocite{Schawinski09} find that AGN host galaxies are transitioning from the blue cloud to the red sequence, supporting such a scenario, whereas Westoby et al. (2007)\nocite{Westoby07} and Cardamone 
et al. (2010)\nocite{Cardamone10} find a large fraction of AGN in passive galaxies. Recently, Symeonidis et al. (2013b)\nocite{Symeonidis13b} found that the overlap in optical colour-colour and colour-magnitude space of star-forming 
galaxies (SFGs) and type II AGN at $z<1.5$ indicates that SFGs are on average three times more likely to host a type II AGN than would be expected 
serendipitously, if AGN and star-formation events were unrelated. 

Although measured star-formation rates (SFRs) of AGN host galaxies span a large range, sometimes reaching a few thousand M$_{\odot}$/yr (e.g. Isaak et al. 2002\nocite{Isaak02}; Priddey et al. 2003a; 2003b
\nocite{Priddey03a}\nocite{Priddey03b}; Lutz et al.2008; Silverman et al. 2009\nocite{Silverman09}; Mainieri et al. 2011\nocite{Mainieri11}, Dai et al. 2012; Khan-Ali et al. 2015; Podigachoski et al. 
2015\nocite{Podigachoski15}), it is not yet clear whether and/or how the AGN accretion rate and its host's SFR are connected. Some studies show that the AGN accretion rate is well correlated with SFR over a 
large range in AGN luminosity (e.g. Rovilos et al. 2012\nocite{Rovilos12}; Chen et al. 2013\nocite{Chen13}; Hickox et al. 2014\nocite{Hickox14}), whereas others report that the correlation is almost non-existent 
at low AGN luminosities but becomes stronger at high AGN luminosities (e.g. Shao et al. 2010\nocite{Shao10}; Rosario et al. 2012\nocite{Rosario12}; 2013\nocite{Rosario13}), or that the two quantities are not 
correlated (e.g. Rawlings et al. 2013; Azadi et al. 2015\nocite{Azadi15}; Stanley et al. 2015\nocite{Stanley15}). Page et al. (2012)\nocite{Page12} and Barger et al. (2015)\nocite{Barger15} show that the most luminous AGN do not 
reside in the most 
highly star-forming hosts, whereas Harrison et al. (2012)\nocite{Harrison12} present contrasting results and Rafferty et al. (2011\nocite{Rafferty11}) find that luminous AGN are more common in systems with 
high SFRs. 

A possible reason for discrepant results amongst these studies, could be the difficulties encountered when dealing with AGN contamination to the galaxies' spectral energy distributions (SEDs) and SFR estimates. Currently, one of 
the most widely used methods of calculating star-formation rates is with IR/submm continuum data, since IR emission is an excellent tracer of young star-forming/starburst regions, particularly in galaxies with 
high SFRs where the contribution from dust heating from older stars is minimal. However, in AGN host galaxies, in addition to starlight-heated dust emission, the infrared SED also includes a contribution from 
the AGN: UV/optical light from the AGN is expected to be intercepted by dust in its vicinity (i.e. the torus) as well as dust distributed over kpc scales in the host galaxy. Indeed, unless the torus is all enveloping and no light can escape, 
then a scenario whereby UV/optical light from the AGN is not intercepted by the dust over kpc scales in the host galaxy is not realistic. A similar argument was made by Sanders et al. (1989\nocite{Sanders89}) who concluded that AGN 
must contribute to dust heating over kpc scales. As a result, when dealing with dust-reprocessed UV/optical radiation from an AGN, two components need to be taken into account: (i) IR emission from the AGN-heated dust in the torus 
and (ii) IR emission from AGN-heated dust in the host galaxy, the former peaking at shorter wavelengths than the latter. Here we aim to provide an intrinsic AGN SED up to 1000\,$\mu$m, reproducing as far as possible the total AGN 
contribution in the far-
IR/submm. Our motivation is simple: if we are to truly understand the connection between AGN and star-formation then we must be confident of the characteristics of the intrinsic AGN emission. 

The paper is laid out as follows: in section \ref{sec:sample} we describe our criteria for the sample selection and finalise the sample
for this work. In section \ref{sec:SEDs} we describe the fitting of SEDs and produce an intrinsic AGN template. In section
\ref{sec:discussion} we discuss our results and we conclude with section \ref{sec:conclusions}. Throughout, we adopt a concordance cosmology of
H$_0$=70\,km\,s$^{-1}$Mpc$^{-1}$, $\Omega_{\rm M}$=1-$\Omega_{\rm \Lambda}$=0.3.

\section{The sample}
\label{sec:sample}

\subsection{Selection criteria}
\label{sec:selection_criteria}
As outlined in section \ref{sec:introduction}, our study aims
to understand the intrinsic AGN SED, particularly in the infrared. Naturally, the ideal sample for this would be galaxies with no stars so that the total broadband SED would be the result of AGN emission only 
(direct or reprocessed). This is of course impossible, so the next best approach is to select a sample of galaxies whose stellar output is small compared to the AGN emission. 
Low luminosity AGN are not appropriate for such a study, as their radiative output is often drowned by their host's over most of the electromagnetic spectrum making their intrinsic SEDs difficult to extract. Our 
first requirement is, therefore, that our sample should be composed of optically-luminous, broad-line AGN which dominate the global SED, at least from X-ray to near-IR wavelengths. The second requirement is 
that the sample is at low redshift where on average star-formation rates are low, but also so that available photometry can probe longer rest-frame wavelengths. Since our aim is to extract the AGN SED from 
the galaxies' total emission, the third requirement is that we have a measure of the SFR independently of the galaxies' broad-band SED. For optically luminous AGN, the optical spectra cannot be used to extract 
information about the host, whereas the mid-IR spectra, and in particular Polycyclic Aromatic Hydrocarbon (PAH) features, are a viable alternative.

\begin{figure}
\epsfig{file=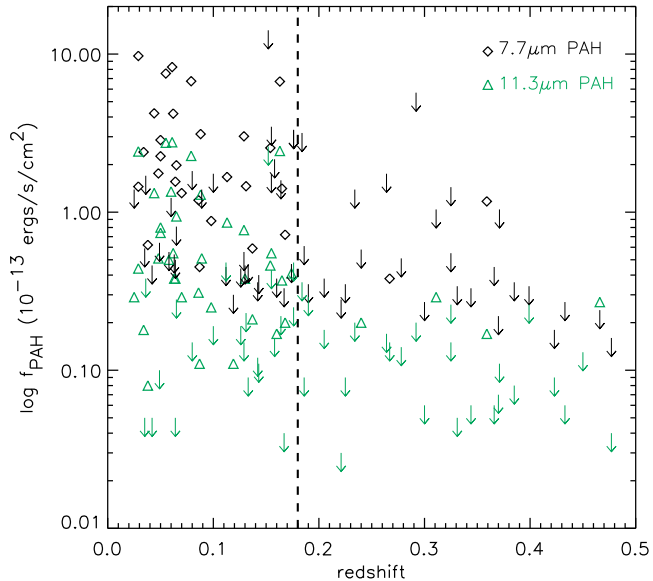, width=0.99\linewidth} 
\caption{PAH flux versus redshift for the PG QSOs from Shi et
  al. (2007). Black diamonds and black upper limits are for the
  7.7$\mu$m PAH feature. Blue triangles and blue upper limits are for
  the 11.3$\mu$m PAH feature. Note that above z=0.18 (dashed line) the PAH detections are severely incomplete - this
  serves as a cut for our sample.}
\label{fig:fPAHz}
\end{figure}

In line with the criteria outlined above, our sample consists of Palomar-Green (PG) QSOs with mid-IR spectroscopy, drawn from Shi et al. (2007). PG QSOs are selected in the $B$ band, have blue $U-B$ colours, a star-like
appearance and broad emission lines; see Schmidt $\&$ Green (1983\nocite{SG83}). Shi et al. (2007) present mid-IR \textit{Spitzer}/IRS spectra and examine the mid-IR
and star-forming properties for the entire parent PG QSO sample up to z=0.5 (see also Schweitzer et al. 2006\nocite{Schweitzer06}; Netzer et al. 2007\nocite{Netzer07}; Veilleux et al. 2009\nocite{Veilleux09}).  
Since the availability of PAH flux measurements is central to our work, in order to determine the redshift range for our sample we examine the distribution of PAH fluxes as a function of redshift (Fig. 
\ref{fig:fPAHz}). It is clear that above $z$=0.18 the PAH detections are sparse, hence we subsequently select PG QSOs at $z<0.18$ (58 sources).

\begin{figure}
\epsfig{file=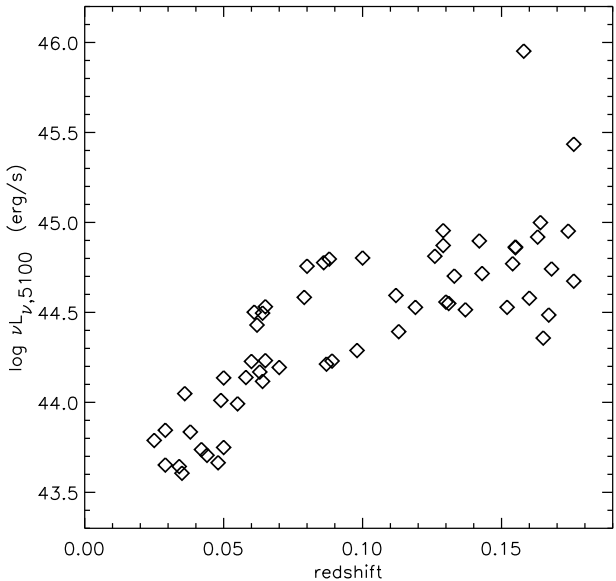,width=0.99\linewidth} 
\caption{Rest-frame luminosity at 5100\AA \,($\nu L_{\nu, 5100}$, erg/s) as a function of redshift for the sample of 58 PG QSOs at $z<0.18$.}
\label{fig:L5100z}
\end{figure}

\subsection{PAH-derived star-formation rates}
\label{sec:PAHSFRs}

PAH features appearing in the mid-IR 6-13$\mu$m (e.g. Allamandola et al. 1985\nocite{Allamandola85}; 1989\nocite{Allamandola89}; Laurent et al. 
1999\nocite{Laurent99}, Sturm et al. 2000\nocite{Sturm00}) have been detected in a wide range of sources and used as diagnostic tools for physical processes in the ISM; they are thought to originate in photo-dissociation regions and 
are excited by UV photons from massive young stars (e.g. Roche $\&$ Aitken 1985\nocite{RA85}; Roche et al. 1991\nocite{Roche91}; Li $\&$ Draine 2002\nocite{LD02}). As a result PAHs are extensively used as star-formation tracers 
(e.g. Peeters et
al. 2004\nocite{Peeters04}; Forster-Schreiber et al. 2004\nocite{Forster-Schreiber04}; Risaliti et al. 2006\nocite{Risaliti06}; Pope et al. 2008\nocite{Pope08}) and although the properties of the PAH spectrum 
vary over small scales and are affected by factors such as metallicity and radiation field hardness (e.g. Roche et al. 1991\nocite{Roche91}; Engelbracht et al. 2008\nocite{Engelbracht08}; Gordon et al. 2008\nocite{Gordon08}; Bendo et 
al. 
2008\nocite{Bendo08}) they are good gauges of star-formation on galactic scales (Roussel et al. 2001\nocite{Roussel01}; Kennicutt et al. 2009\nocite{Kennicutt09}; Treyer et al. 2010\nocite{Treyer10}) and have been seen to correlate 
well with other star-formation tracers such as H$\alpha$ (e.g Kennicutt et al. 2009), [Ne II] (e.g. Ho $\&$ Keto 2007; Shipley et al. 2013) and the total infrared luminosity ($L_{\rm IR}$; e.g. Peeters et al. 2004; Brandl et al. 
2006\nocite{Brandl06}, Hanami et al. 2012\nocite{Hanami12}). Although PAHs are prevalent in star-forming galaxies, early studies of AGN mid-IR spectra reported a lack of PAH features around some AGN (e.g. Cutri et al. 1984; Aitken 
$\&$ Roche 1985; Roche et al. 1991), which led to the idea that PAHs might be destroyed by hard radiation from the AGN (e.g. Aitken $\&$ Roche 1985; Voit 1992). However, subsequent studies have shown that PAHs are common 
features in galaxies which host AGN and as such they have been routinely employed as star-formation tracers in AGN hosts (e.g. Shi et al. 2007\nocite{Shi07}; Lutz et al. 2008; Watabe et al. 2008\nocite{Watabe08}; Rawlings et al. 
2013\nocite{Rawlings13}; Esquej et al. 2014\nocite{Esquej14}).  

Shi et al. measure the strength of the 7.7 and 11.3$\mu$m PAH emission features in the PG QSO spectra used in this work and derive the conversion factor between PAH luminosity and 8-1000$\mu$m total infrared luminosity 
attributed to star-formation ($L_{\rm SFIR}$) for each PG QSO, by adopting an SED template from the work of Dale et al. (2001\nocite{Dale01}) and Dale $\&$ Helou (2002\nocite{DH02}) that gives the closest PAH emission line flux at 
the redshift of the object; see Shi et al. (2007) for more details. We note that the $L_{\rm SFIR}$/$L_{\rm PAH}$ conversions in Shi et al. are consistent with other works, e.g. Lutz et al. (2003\nocite{Lutz03}); Smith et al. 
(2007\nocite{Smith07}); Hern{\'a}n-Caballero et al. (2009\nocite{Hernan-Caballero09}). In cases where both 7.7 and 11.3$\mu$m PAH features are detected, Shi et al. (2007) use the 11.3$\mu$m PAH to derive $L_{\rm SFIR}$ as its 
uncertainty is lower than that of the 7.7$\mu$m PAH and it is thought to be less susceptible to suppression by the AGN radiation (e.g. see Smith et al. 2007\nocite{Smith07}; O'Dowd et al. 2009\nocite{ODowd09}; Hunt et al. 
2010\nocite{Hunt10}; Diamond-Stanic $\&$ Rieke 2010\nocite{DSR10}, Esquej et al. 2014,\nocite{Esquej14} Shi et al. 2014\nocite{Shi14}). In our final sample (section 
\ref{sec:final_sample}), there is only one source for which $L_{\rm SFIR}$ is derived using the 7.7$\mu$m PAH instead of the 11.3$\mu$m one, as the latter is not detected (in section \ref{PAHdestr} we examine the effect this has on 
our results).

\begin{figure}
\epsfig{file=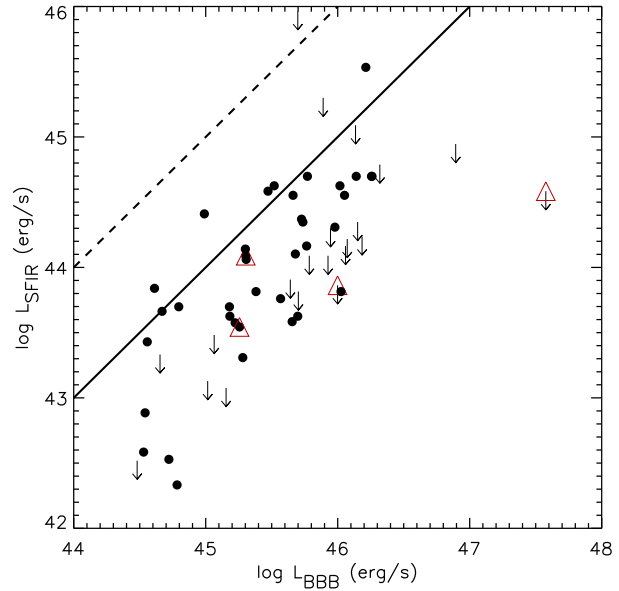,width=0.99\linewidth} 
\caption{Total infrared luminosity attributed to star-formation ($L_{\rm SFIR}$; 8-1000$\mu$m) vs AGN big blue bump luminosity ($L_{\rm BBB}$; 2keV--1$\mu$m) for the 58 PG QSOs at
  $z<$0.18. The dashed line is the one-to-one relation and the solid line
represents $L_{\rm BBB} =10 \times L_{\rm SFIR}$. Red triangles indicate the
radio-loud sources. Our final sample
includes all objects below the solid line apart from the ones that are identified as radio-loud. }
\label{fig:L_BBB_LSFIR}
\end{figure}

\subsection{Photometric data}
We obtain archival and published photometry for the 58 $z<0.18$ PG QSOs as follows: Palomar $B$-band photometry
(Shi et al. 2014\nocite{Shi14}; see also Schmidt $\&$ Green 1983), SDSS magnitudes
from the SDSS Photometric Catalog, Release 9 (Ahn et al. 2012\nocite{Ahn12}), 2MASS magnitudes from the 2MASS All-Sky Catalog of Point
Sources (Skrutskie et al. 2006\nocite{Skrutskie06}), \textit{WISE} magnitudes  from the AllWISE Source
Catalog\footnote{http://wise2.ipac.caltech.edu/docs/release/allwise/},
\textit{Spitzer}/MIPS data from Shi et al. (2014), ISO flux densities
(Haas et al. 2000\nocite{Haas00}; 2003\nocite{Haas03}), \textit{IRAS} photometry from the \textit{IRAS} point source and faint source
catalogues, \textit{AKARI} 9 and 18$\mu$m data from the \textit{AKARI}/IRC mid-IR all-sky
Survey and \textit{AKARI}/FIS All-Sky Survey Point Source fluxes at  65, 90,
140, 160$\mu$m, \textit{Herschel} (Pilbratt et al. 2010\nocite{Pilbratt10})\footnote{Herschel is an ESA space observatory with science instruments provided by European-led Principal Investigator consortia and with important 
participation from NASA}/PACS (Poglitsch et al. 2010\nocite{Poglitsch10}) 160$\mu$m flux densities from
Shi et al. (2014) and radio data from Haas et al. (2003). The \textit{Herschel}/SPIRE (Griffin et al. 2010\nocite{Griffin10}) data are from the \textit{Herschel} Science Archive, processed by the SPIRE HIPE pipeline version 12 (Ott 
2010\nocite{Ott10}).
The photometry was extracted at the source position using the SUSSEXTractor task within HIPE (Savage $\&$ Oliver 2007\nocite{SO07}, Smith et al 2012\nocite{SmithAJ12}, Pearson et al 
2014\nocite{Pearson14}). The SUSSEXtractor task estimates the flux density from an image smoothed with a convolution kernel derived from the SPIRE beam FWHM. In the cases where our targets  were 
observed in the SPIRE small map mode, the flux density measured by SUSSEXtractor was verified using the SPIRE Timeline Fitter (Bendo et al. 2013\nocite{Bendo13}) which fits a  two dimensional elliptical  
Gaussian function at the source position in the  timeline data. The agreement between the SUSSEXtractor and Timeline Fitter flux densities was found to be better than $\sim$5 per cent. 

The detection statistics are as follows: 4 sources have no significant detections above 25$\mu$m ($\sim$7 per cent of the sample) and 5 sources have no significant detections above 100$
\mu$m ($\sim$9 per cent). These 9 sources also have upper limits in $L_{\rm SFIR}$; in total there are 21 sources with upper limits in $L_{\rm SFIR}$ ($\sim$36 per cent of the sample).

We calculate rest-frame 5100\AA \,luminosities ($\nu L_{\nu}$), by interpolating between
the SDSS bands. For the 9 objects which do not
have SDSS counterparts we interpolate between the Palomar $B$-band and the 2MASS $J$-band. To convert to big blue bump
luminosities ($L_{\rm BBB}$; 2keV--1$\mu$m - Grupe et
al. 2010), we use the Grupe et al. (2010) expression, $\rm log$ $L_{\rm BBB} = (1.32 \pm 0.06)
\times$ $\rm log$ $\nu L_{\nu, 5100} - (10.84 \pm 2.21)$. The range in $\nu L_{\nu, 5100}$ covered by our sample is three orders of magnitude (10$^{43}$ to 10$^{46}$\, erg/s; see Fig \ref{fig:L5100z}), 
translating to log\,[$L_{\rm BBB}$ (erg/s)]=44.5-48, and is a
strong function of redshift. According to the local AGN optical luminosity function of Schulze et al. (2009\nocite{Schulze09}), our sample covers the whole luminosity range above
$L_{\star}$, hence corresponds to the most luminous nearby AGN.

\subsection{The final sample}
\label{sec:final_sample}
Figure \ref{fig:L_BBB_LSFIR} shows $L_{\rm SFIR}$ (the total infrared luminosity attributed to star-formation) as a function of
$L_{\rm BBB}$. Note that all bar one source have $L_{\rm BBB}> L_{\rm SFIR}$
and for the majority $L_{\rm BBB}> 10 \times L_{\rm SFIR}$. This confirms the
suitability of this sample for our study, i.e. sources where the AGN
is more powerful than the galaxies' stellar output. For our analysis we keep the sources with $L_{\rm BBB}> 10 \times L_{\rm SFIR}$, where the AGN is an order of
magnitude more luminous than the luminosity from star-formation. In this way we maximise the ratio of AGN to stellar powered emission. 

We remove radio-loud AGN from our final sample as their far-IR emission is contaminated by non-thermal
processes. Kellermann et al. (1989\nocite{Kellermann89}) examine the radio-loudness of the
PG sample, using a radio loudness criterion ($R$) defined as the
ratio of radio flux at 6\,cm to optical flux at 4400$\AA$. We take $R$=10 as
the dividing line between radio-quiet and radio-loud AGN. 4 sources are
identified as radio-loud. 

The final sample consists of 47, radio-quiet PG QSOs at $z<0.18$, with
$L_{\rm BBB}> 10 \times L_{\rm SFIR}$. Hereafter our analysis involves only the final sample.

\section{Method and Results}
\label{sec:SEDs}

\subsection{Fitting QSO SEDs}
\label{sec:SED_fitting}

We do the SED fitting as follows: from 22$\mu$m onwards (just after the 18$\mu$m
silicate feature) we fit with a
power-law/greybody combination, whereas up to 22$\mu$m we interpolate logarithmically between the bands. The fitting includes the greybody (GB) function for the far-IR to describe the emission from large 
grains in equilibrium, and a power-law
(PL) for the mid-IR to approximate the emission from hot dust, combined at a critical frequency $\nu_{\ast}$  (see also Blain
et al. 2003\nocite{BBC03}; Younger et al. 2009\nocite{Younger09}) as follows:
\begin{equation}
F_{\nu} \propto  \left\{
  \begin{array}{l l}
    \frac{\nu ^{3+\beta}}{e^{(h\nu/kT_{\rm dust})} - 1} & \quad $if$ \quad \nu < \nu_{\ast}\\
    \nu^{\alpha} & \quad  $if$ \quad \nu > \nu_{\ast}
  \end{array} \right.\
\end{equation}
where $F_{\nu} $ is the flux density, $h$ is the Planck constant, $c$ is the speed of light in a
vacuum, $k$ is the Boltzmann constant, $T_{\rm dust}$ is the temperature of the
greybody function and $\beta$ is the emissivity --- we adopt $\beta$=1.5, consistent with studies of the far-IR emissivity of large
grains (Desert, Boulanger $\&$ Puget 1990\nocite{DBP90}; see also Boselli et al. 2012\nocite{Boselli12}). At the critical
frequency $\nu_{\ast}$ the slopes of the two functions are equal
and hence $\alpha$=$d$log GB/$d$log$\nu$.
We perform  $\chi ^2$ fitting in order to find the best fit GB/PL
combination, with $\alpha$, $\nu_{\ast}$ and $T_{\rm dust}$ as free parameters. 
The SEDs for all 47 sources in our final sample are shown in Fig. \ref{fig:SEDs_QSO_average} normalised at 20$\mu$m, and also individually in Appendix A. The small peak at 20$\mu$m is the 
silicate feature in emission (see Shi et al. 2014). This is quite common in unobscured QSOs, as the emission from silicates in the inner part of the torus is visible (e.g. Siebenmorgen et al. 
2005\nocite{Siebenmorgen05}, Sturm et al. 2005\nocite{Sturm05}, Shi et al. 2006\nocite{Shi06}). Note that our SEDs can be considered well-sampled up to about 500$\mu$m (rest-frame).

\begin{figure}
\epsfig{file=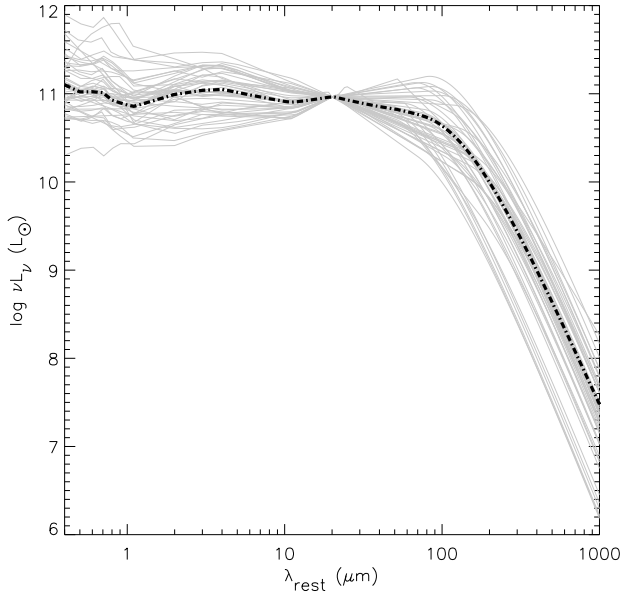,width=0.99\linewidth} 
\caption{The average SED of the sample of 47 QSOs (black dot-dashed curve). The average SED is plotted in actual luminosity units, whereas the individual QSO SEDs (grey curves) are normalised to the average SED at 20$\mu$m.}
\label{fig:SEDs_QSO_average}
\end{figure}

\begin{figure}
\epsfig{file=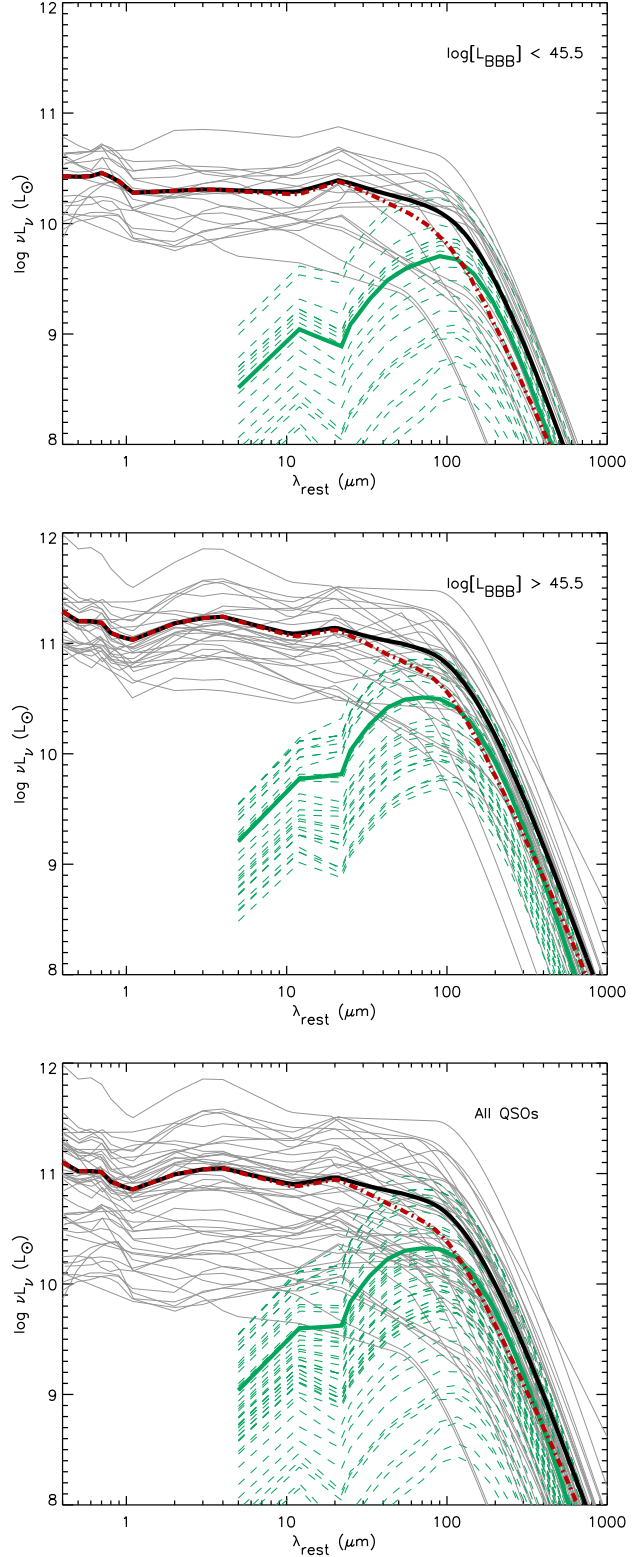,width=0.99\linewidth} 
\caption{SEDs for the PG QSOs in our sample (black curves) and their average (thick black curve), as well as the SF templates (green dashed curves) and their average (thick green curve). 
The intrinsic AGN SED is shown by the red dot-dashed curve and is obtained by subtracting the thick green curve from the thick black curve. In the first two panels the sample is split into two AGN luminosity bins: log\,$[L_{\rm
    BBB}]<45.5$ and log\,$[L_{\rm
    BBB}]>45.5$ and the last panel shows all QSOs together. }
\label{fig:lumQSO_SED_average_wDH02}
\end{figure}

\subsection{Matching star-forming templates to QSOs}
\label{sec:matching}

We now assign a star-forming SED template to each PG QSO. As these are
low redshift systems, dust temperature evolution is not significantly
at play (see Symeonidis et al. 2013a\nocite{Symeonidis13a}), hence it is appropriate to use SED templates for nearby
galaxies. To be consistent with how the $L_{\rm SFIR}$ values were derived in Shi et al. (2007; see section \ref{sec:PAHSFRs}), we use the Dale $\&$ Helou (2002; hereafter DH02) SED library. As in Shi et al., we assign total infrared 
luminosities ($L_{\rm IR}$; 8--1000$\mu$m) to the DH02 SED library using their 60/100$\mu$m colour and the relation in Marcillac et al. (2006). Each SED template is subsequently normalised to units of L$_{\odot}$ by its $L_{\rm IR}
$.
For each PG QSO, we chose the DH02 template whose $L_{\rm IR, templ}$ is closest to the QSO's $L_{\rm SFIR}$ calculated by Shi et al. (2007). This is an important step as there is a correlation between $L_{\rm IR}$ and dust
temperature (or SED peak) (e.g. Dunne et al. 2000\nocite{Dunne00}; Dale et al. 2001\nocite{Dale01}; see also Symeonidis et al. 2013a), so for each PG
QSO we need to have the appropriate far-IR shape of star-forming (SF) emission. 

For each QSO, the corresponding DH02 template is scaled by the factor $L_{\rm SFIR}$/$L_{\rm IR, templ}$. For the 17 sources (out of 47; 36 per cent) which have a 3$\sigma$ upper limit in
$L_{\rm SFIR}$ due to low significance PAH detections, we assume that their $L_{\rm SFIR}$ is equal to half the upper limit and scale the corresponding DH02 templates by the factor $\frac{L_{\rm SFIR, lim}/2}
{L_{\rm IR, templ}}$. This is a reasonable assumption given that they could in principle have any $L_{\rm SFIR}$ from 0 to $L_{\rm SFIR, lim}$; in section \ref{sec:PAHlim} we evaluate the effect that this has 
on our results.

\subsection{The intrinsic AGN SED}
\label{sec:pureAGN}

Our PG QSO SEDs are broadband, hence to subtract the SF DH02 templates which include spectral data in the mid-IR, we resample the 2--30$\mu$m part of the rest-frame DH02 SED into \textit{WISE} filter-convolved luminosities at 
3.4, 4.6, 12 and 22$\mu$m and interpolate between them. The average SF SED is then subtracted from the average PG QSO SED from 5$\mu$m onwards, which is the onset of the dust continuum; the residual we obtain is the 
average \textit{intrinsic} or \textit{pure} AGN SED. We confirm that we get the same intrinsic AGN SED if instead we calculate the residual SED for each PG QSO and subsequently average those. 
Note that our intrinsic AGN SED can be considered well-sampled up to about 500$\mu$m (rest-frame). 

In Appendix \ref{appendixC}, we test the method used to calculate the average star-forming infrared SED for our sample of QSOs (see this section and section \ref{sec:PAHSFRs}), on 2 sets of pure starburst galaxies with known far-IR 
SEDs. We find that the average far-IR SED of each set of starbursts is reproduced correctly, and thus we have confidence that this method accurately predicts the average star-forming SED corresponding to the input PAH 
measurements.  

The matched SF templates and QSO SEDs are shown in Fig. \ref{fig:lumQSO_SED_average_wDH02} split into two luminosity groups each covering about 1.5 dex in $L_{\rm BBB}$ above and below log\,$L_{\rm BBB}$=45.5 and 
then all together. Note that shortward of about 20$\mu$m the average QSO SED is typically at least an order of magnitude brighter than the average SF SED and the AGN entirely dominates the 
broad-band emission, whereas longwards of 20$\mu$m there is an additional contribution from the host galaxy. The lower panel shows that the average luminosity from star-formation is less than the average intrinsic AGN luminosity up 
to $\sim$100\,$\mu$m and approximately equal to it thereafter, suggesting that in some of these sources the whole SED is AGN-dominated. 

\begin{figure}
\epsfig{file=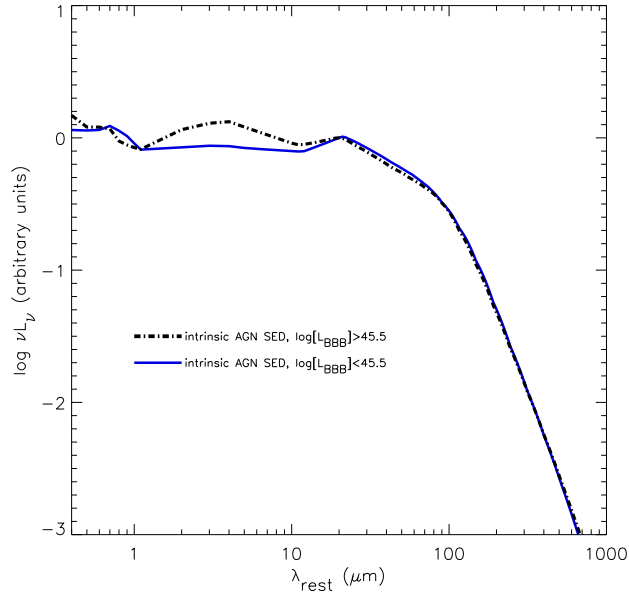,width=0.99\linewidth} 
\caption{The intrinsic AGN SED derived separately from the log\,$[L_{\rm
    BBB}]<45.5$ and log\,$[L_{\rm BBB}]>45.5$ QSOs, normalised at 20$\mu$m.}
\label{fig:SEDs_QSO_average_comp}
\end{figure}

\begin{figure}
\epsfig{file=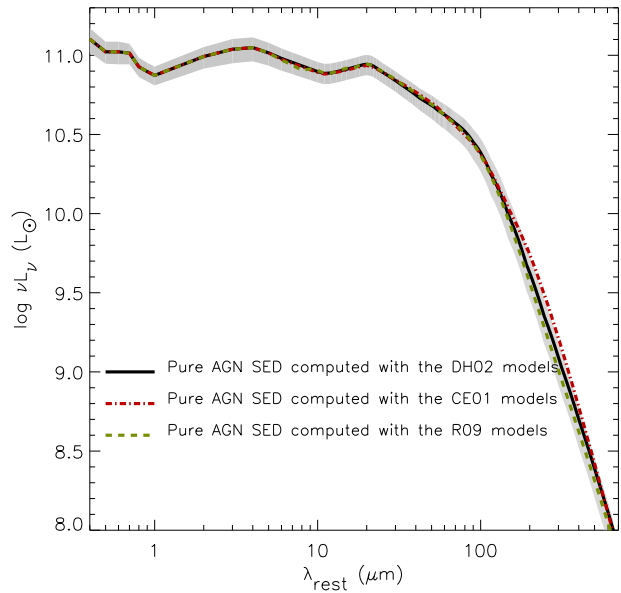,width=0.99\linewidth} 
\caption{The intrinsic AGN SED computed with Dale $\&$ Helou 2002 (DH02) library (solid black curve and 68 per cent confidence intervals shown as a shaded region), compared with the one computed with the Chary $\&$ Elbaz 
2001 (CE01) and Rieke et al. 2009 (R09) SED templates --- see section \ref{sec:pureAGN}. Note that throughout this work we use the one computed with the DH02 library, although the difference is small between the three SEDs.}
\label{fig:QSO_residualSEDs_compareSEDmodels}
\end{figure}

Fig. \ref{fig:SEDs_QSO_average_comp} compares the intrinsic AGN SEDs for the two QSO luminosity bins, log\,$\nu L_{\nu, BBB}<45.5$ and log\,$\nu L_{\nu, BBB}>45.5$ normalised at 20$\mu$m, where the 
SEDs change from AGN dominated to AGN+host. We note that there is a small difference in the 1-10$\mu$m region between the low luminosity and high luminosity QSOs of about 0.2\,dex, with the more 
luminous sources having a more pronounced 1-10$\mu$m bump in that wavelength range.  Such a mid-IR bump has been observed before in luminous AGN; e.g. Leipski et al. (2014\nocite{Leipski14}) find that 
a separate hot blackbody component ($T\sim$1200\,K) is needed to fit the 1-10$\mu$m SED of their QSO sample. Emission from the accretion disk is not significant past 1$\mu$m (e.g. Laor 
1990\nocite{Laor90}; Collinson et al. 2015\nocite{Collinson15}), so this feature is likely caused by scattered and reprocessed emission by dust (e.g. see also models by Fritz et al. 2006\nocite{Fritz06}). 
In contrast to the mid-IR, there is negligible difference in the intrinsic  SEDs of the lower and higher luminosity AGN longward of 20$\mu$m. In the remainder of 
this work we use the intrinsic AGN SED derived from the whole sample (see lower panel of Fig. \ref{fig:lumQSO_SED_average_wDH02}) and publish this in table \ref{table:residual_sed}.

To calculate the 68 per cent confidence intervals on the intrinsic AGN SED we bootstrap the residual SEDs of our sample of 47 QSOs and hence re-derive an intrinsic AGN SED 5000 times. Since $L_{\rm SFIR}$ is used to assign a SF 
template, which is in turn used to compute the residual SEDs, the uncertainties on $L_{\rm SFIR}$ (which are a combination of the $L_{\rm SFIR}$/$L_{\rm PAH}$ conversion factor uncertainties and PAH flux measurement errors) are propagated to the residual SEDs. The distribution of \textit{observed} residual SEDs is thus the result of the convolution of the \textit{intrinsic} distribution of residual SEDs and the error distribution and is consequently broader than the 
intrinsic distribution. Since the bootstrapping is performed on the observed distribution of residual SEDs, the resulting error on the mean has all random errors folded in. We find that the 1$\sigma$ error on the pure AGN SED ranges between 12 and 45 per cent as a function of wavelength (see Fig. \ref{fig:AGNSED}). 

\subsection{Effect of PAH upper limits on the intrinsic AGN SED}
\label{sec:PAHlim}
We remind the reader that for the sources which have low significance PAH detections, we have taken $L_{\rm SFIR}$=$L_{\rm SFIR, lim}$/2 (see section \ref{sec:matching}). We now examine how the intrinsic AGN SED changes if 
we set $L_{\rm SFIR}$ to the minimum and maximum values, i.e. $L_{\rm SFIR}$=0 and $L_{\rm SFIR}$=$L_{\rm SFIR, lim}$. We find that the pure AGN SED changes by up to 30 per cent in the 20--1000$\mu$m range, within the 
68 per cent confidence intervals we obtain from bootstrapping.

\subsection{Effect of SED library choice on the intrinsic AGN SED}
We assess the effect of our choice of star-forming SED library, by re-computing the intrinsic AGN SED with two different star-forming SED libraries, the Rieke et al. (2009\nocite{Rieke09}) templates and the Chary $\&$ Elbaz 
(2001\nocite{CE01}) templates (note that the Rieke et al. library does not extend to log\,[$L_{\rm IR}$/L$_{\odot}$]$<$9.75, so we assign the log\,[$L_{\rm IR}$/L$_{\odot}$]$<$9.75 template to all our $L_{\rm SFIR}<10^{10}$\,L
$_{\odot}$ sources). Fig. \ref{fig:QSO_residualSEDs_compareSEDmodels} shows that there is small difference in the resulting intrinsic AGN SED.

\section{Discussion}
\label{sec:discussion}

We found that the average intrinsic AGN emission in a sample of nearby QSOs dominates the broadband SED up to about 100$\mu$m and is comparable to the average star-forming emission thereafter. Before we examine the implications of our results in sections \ref{sec:otherAGNSEDs} and \ref{sec:implications}, we dedicate section \ref{sec:considerations} to a discussion regarding the star-forming  luminosities in our sample of PG QSOs.

\subsection{Considerations on the star-forming luminosities of QSO host galaxies}
\label{sec:considerations}

Recently, Petric et al. (2015\nocite{Petric15}) reported their results
on the far-IR emission of PG QSOs, finding an offset between IR
SED-derived SFRs and PAH-derived SFRs, with the former being
systematically larger. This result is consistent with ours, i.e. that
the star-formation implied by PAHs does not account for all the far-IR
emission in nearby QSOs. Petric et al. (2015) suggest
that as well as the heating of dust by the AGN, heating of the dust by
old stars, and suppression of the PAHs by the AGN may be important.

\subsubsection{Could evolved stars contribute to dust heating?}
In some nearby low-luminosity, low-SFR galaxies a fraction of the
far-IR emission is heated by the evolved stellar population rather
than young stars (e.g. Bendo et al. 2010\nocite{Bendo10};
2012\nocite{Bendo12}; 2015\nocite{Bendo15}). This suggests that for
the 8 QSOs in our sample with measured $L_{\rm SFIR}<10^{10}
\,L_{\odot}$, scaling the templates according to $L_{\rm SFIR}$ might
underestimate the normalisation of the template with respect to the
QSO SED. Nevertheless, the $L_{\rm SFIR}$ of these QSOs is very low
compared to the average $L_{\rm SFIR}$ of the sample, thus even
increasing the normalisation by a factor of 3 (corresponding to an
extreme case whereby only 30 per cent of the total IR luminosity is
due to dust heated by star-formation; see Bendo et al. 2015) has a
negligible effect on our intrinsic AGN SED.

\subsubsection{How well do PAHs trace star-formation in AGN host galaxies?}
\label{PAHdestr}
PAH molecules are not expected (or observed) to survive in the
vicinity of the AGN, where they are directly illuminated by its hard
radiation field (Siebenmorgen, Kr\"ugel \& Spoon
2004\nocite{SKS04}). However, there has been a long-standing
debate about whether PAHs are suppressed further out in the host
galaxies of AGN, where the ultraviolet radiation field is dominated by
stars.  In theory X-rays from the AGN are expected to destroy PAHs as
a consequence of photoionization, the relevant transition being the
carbon K edge at 0.288 keV in the soft X-rays (Voit 1992). The PAH
survival condition suggested by Voit (1992) is that PAHs containing
$\gtrsim$50 carbon atoms can survive at $\sim$1\,kpc from an AGN of X-ray
luminosity $\nu L_{\nu}$=10$^{44}$erg\,s$^{-1}$ if they are shielded from
the AGN by column densities $>10^{22}$~cm$^{-2}$. Typically, AGN tori have column
densities more than an order of
magnitude larger than this (e.g. Risaliti, Maiolino \& Salvati
1999\nocite{RMS99}; Buchner et al. 2014\nocite{Buchner14}) and their covering factors are 50 per cent or more (e.g. Maiolino $\&$ Rieke 1995\nocite{MR95}, Dwelly $\&$ Page 2006\nocite{DP06}). In addition, observational evidence suggests that AGN tori are aligned with the galaxy plane (e.g. Maiolino $\&$ Rieke 1995\nocite{MR95}; Lagos et al. 2011\nocite{Lagos11}). As a result, PAHs in star-forming regions are likely to be well
shielded by the torus along the majority of lines of sight in the galaxy. 

Some studies argue that the smaller PAH molecules responsible for the shorter wavelength PAH features might be suppressed in the host galaxies of AGN, but there is strong empirical evidence that this is not the case for the 11.3\,$\mu$m feature. For example, Smith et
al. (2007\nocite{Smith07}) find that in their {\em Spitzer} study of 59 nearby
galaxies, some that host AGN have lower ratios of
7.7~$\mu$m to 11.3~$\mu$m PAH emission than those that do not, and
with a more pronounced trend closer to the nuclei of the
galaxies. O'Dowd et al. (2009\nocite{Odowd09}) show a similar trend
of reduced 7.7\,$\mu$m PAH emission in some of the AGN-hosting members
of their sample of 92 $z\sim0.1$ galaxies. They argue that the
molecules responsible for the 7.7~$\mu$m emission are likely to be
susceptible to destruction by X-rays or shocks from the AGN, but that
the 11.3~$\mu$m emission is likely to come from larger molecules that
are more resilient. Subsequently, Diamond-Stanic \& Rieke (2010\nocite{DSR10}) studied a sample of 35 Seyfert galaxies and
showed that while the shorter wavelength PAH features are suppressed
in some of these objects, the 11.3~$\mu$m emission obeys the same
relationship with [Ne II] as purely star-forming objects, implying
that it is a robust tracer of star formation rate regardless of the
presence of an AGN. Finally Shipley et al. (2013) conclude that the AGN do not affect PAHs on galaxy-wide scales based on the lack of any trend between the $L_{\rm PAH (7.7)}$/$L_{\rm PAH (11.3)}$ ratio and the AGN luminosity or hardness of the radiation field.

\begin{figure}
\epsfig{file=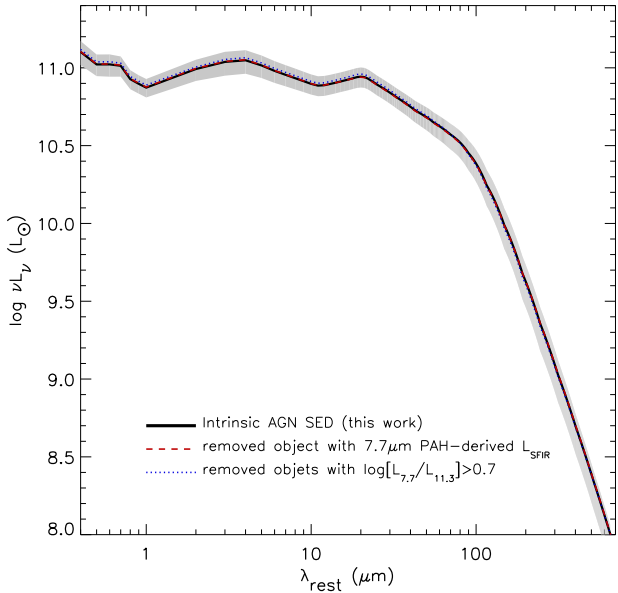,width=0.99\linewidth} 
\caption{The intrinsic AGN SED (solid black curve and 68 per cent confidence intervals shown as a shaded region), re-computed without the object with a 7.7\,$\mu$m PAH-derived $L_{\rm SFIR}$ (red dashed curve) and without the 5 objects with elevated $L_{\rm PAH (7.7)}$/$L_{\rm PAH (11.3)}$ ratios (blue dotted curve); note that the 3 curves are almost indistinguishable.}
\label{fig:AGNrmobj}
\end{figure}

Only one object (PG 0838+770) in our sample of nearby
PG QSOs has $L_{\rm SFIR}$ value based on the 7.7\,$\mu$m PAH rather than the 11.3\,$\mu$m feature. Noting the potential suppression of 7.7\,$\mu$m PAH features by AGN, we have
checked the extent to which the inclusion of this object affects our
intrinsic AGN SED by re-computing the SED with this object excluded
from the sample. We find that the intrinsic AGN changes by less than 1.4 per cent; see Fig. \ref{fig:AGNrmobj}.

\begin{figure}
\epsfig{file=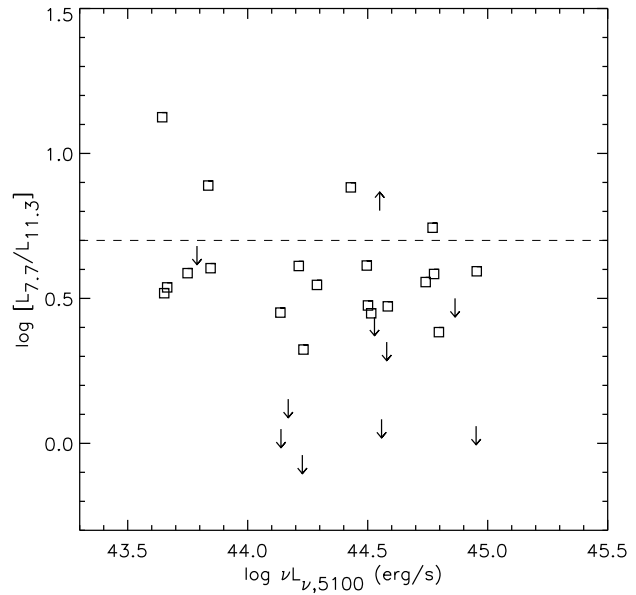,width=0.99\linewidth} 
\caption{Plot of the 7.7/11.3$\mu$m ratio as a function of AGN luminosity (at 5100\AA). Only objects for which at least one of the features is detected are shown. Squares are objects with both features detected, upper limits are for the objects for which the 7.7$\mu$m feature is not detected and the lower limit is for the object with undetected 11.3$\mu$m PAH. Above the dashed line at log\,[$L_{\rm PAH (7.7)}$/$L_{\rm PAH (11.3)}$]$>0.7$ we consider the QSOs to have elevated $L_{\rm PAH (7.7)}$/$L_{\rm PAH (11.3)}$ ratios, higher than the typical values seen in star-forming galaxies.}
\label{fig:LPAHLAGN}
\end{figure}

Ground-based studies using 8--10\,m telescopes have allowed the PAH
emission from the host galaxies of nearby AGN to be studied at much
higher spatial resolution than the space-based studies. Recently,
Esquej et al. (2014\nocite{Esquej14}) examined the PAH emission from a
sample of 29 Seyfert galaxies and found no evidence for the
suppression of the 11.3\,$\mu$m emission even within a few tens of
pc from the nucleus. Alonso-Herrero et al. (2014\nocite{Alonso-Herrero14}) find in their study of six
local AGN that 11.3\,$\mu$m emission is seen at 10\,pc from the
nucleus, suggesting that the molecules responsible for this emission feature
are robust to AGN destruction even in the innermost regions of
the host galaxy. In a follow up case study of the nearby Seyfert
Mrk\,1066, Ramos-Almeida et al. (2014\nocite{Ramos-Almeida14b})
explicitly calculate that when the nuclear continuum is subtracted,
the equivalent width of the 11.3$\mu$m PAH emission in the inner
region of this galaxy is consistent with that seen in star-forming
regions, supporting the validity of the 11.3\,$\mu$m emission as a
tracer of the star-formation rate in AGN host galaxies.

On the other hand, one study has raised concerns about the
use of the 11.3\,$\mu$m feature in AGN hosts. LaMassa
et al. (2012\nocite{LaMassa12}) find that a number of sources in their sample of
Seyfert 2 galaxies show abnormally large ratios of 7.7 $\mu$m to
11.3~$\mu$m PAH emission. LaMassa et al. (2012) interpret these ratios as
evidence for suppression of the 11.3 $\mu$m PAH emission by the AGN, though 
extinction offers an alternative explanation for large 11.3\,$\mu$m / 7.7\,$\mu$m PAH ratios, as noted by Diamond-Stanic \& Rieke (2010\nocite{DSR10}). We compare the distribution of the $L_{\rm PAH (7.7)}$/$L_{\rm PAH (11.3)}$ ratio in our sample of PG QSOs with that of the local star-forming galaxy samples of 
Brandl et al. (2006); LaMassa et al. (2012); Stierwalt et al. (2014), and find them to be broadly consistent with only 5 PG QSOs having elevated $L_{\rm PAH (7.7)}$/$L_{\rm PAH (11.3)}$ ratios, higher than the typical values 
seen in star-forming galaxies (i.e. log\,[$L_{\rm PAH (7.7)}$/$L_{\rm PAH (11.3)}$]$>0.7$). We check the extent to which inclusion of the 5 objects with elevated $L_{\rm PAH (7.7)}$/$L_{\rm PAH (11.3)}$ (PG\,0838+770, PG\,1115+407, PG\,1310-108, PG\,1535+547, PG\,2130+099) affects our results, by re-computing the intrinsic AGN SED with these objects excluded. The resulting SED differs at most by 5.1 per cent from that derived from the whole sample; see Fig \ref{fig:AGNrmobj}. Finally we inspect the 7.7\,$\mu$m/11.3\,$\mu$m PAH luminosity ratio as a function of AGN luminosity (at 5100\AA) for our sample of QSOs (Figure \ref{fig:LPAHLAGN}; note only objects with at least one PAH detection are shown). We see no trend between the $L_{\rm PAH (7.7)}$/$L_{\rm PAH (11.3)}$ ratio and AGN luminosity for the PG QSOs with elevated $L_{\rm PAH (7.7)}$/$L_{\rm PAH (11.3)}$. Whatever the reason is for the elevated 7.7\,$\mu$m/11.3\,$\mu$m PAH ratios in these 5 objects, they do not have a significant impact on the results.

\begin{figure}
\epsfig{file=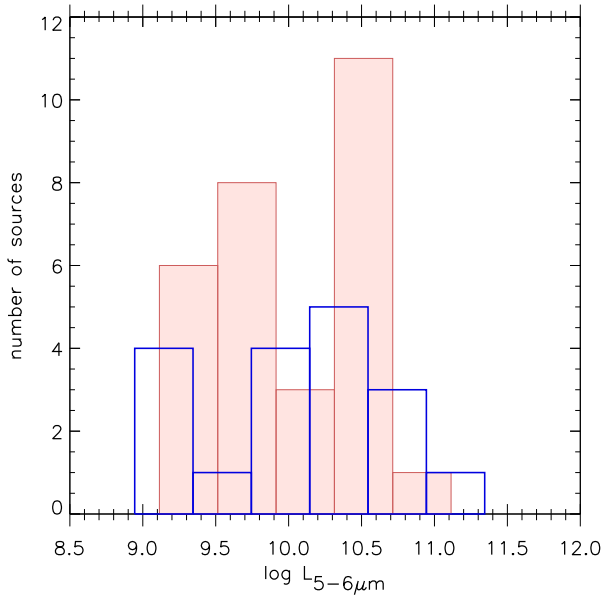,width=0.99\linewidth} 
\caption{The distribution in QSO luminosity (integrated in the 5-6$\mu$m region; values taken from Shi et al. 2007) for our sample, split into sources which have PAH detections (red filled histogram) and sources which have no PAH detections (blue histogram).}
\label{fig:petricplot}
\end{figure}

We also investigate the concern raised by Petric et al. (2015), that the 11.3 PAH detection rate is lower for the most IR-luminous AGN, which might arise as a result of AGN suppression of the 11.3$\mu$m PAH emission. For our sample of QSOs, we show in Fig. \ref{fig:petricplot} the distributions in $L_{5-6\mu m}$ luminosity (i.e. the integrated luminosity in the 5--6$\mu$m range), for sources which have 11.3$\mu$m PAH detections and those that do not. The values of $L_{5-6\mu m}$ were taken from Shi et al. (2007). We find no offset in the two distributions, indicating that the 11.3$\mu$m PAH detection rate is not dependent on AGN luminosity.

The SEDs constructed separately for the QSOs with log\,[$L_{\rm BBB}$]
above and below 45.5 (Figs \ref{fig:lumQSO_SED_average_wDH02} and \ref{fig:SEDs_QSO_average_comp}), offer another opportunity to examine the
possibility that the 11.3\,$\mu$m PAH emission is suppressed by the
AGN. If the AGN were suppressing the 11.3\,$\mu$m emission, we would
expect a larger effect in the higher luminosity subset: the stronger
AGN radiation field would produce stronger suppression of the
11.3\,$\mu$m emission, so that the SFR would be systematically
underestimated by a larger degree in the more luminous QSOs, which
would translate to a larger flux at long wavelengths ($>70$\,$\mu$m) in
the intrinsic AGN SED for the more-luminous QSOs than for the
less-luminous QSOs. Instead, the shapes of the intrinsic AGN SEDs for
the higher and lower luminosity subsamples are almost
indistinguishable beyond 20$\mu$m
(Fig. \ref{fig:SEDs_QSO_average_comp}), and hence no evidence for the
suppression of 11.3~$\mu$m PAH emission by the AGN can be found in the
comparison.

\begin{figure*}
\epsfig{file=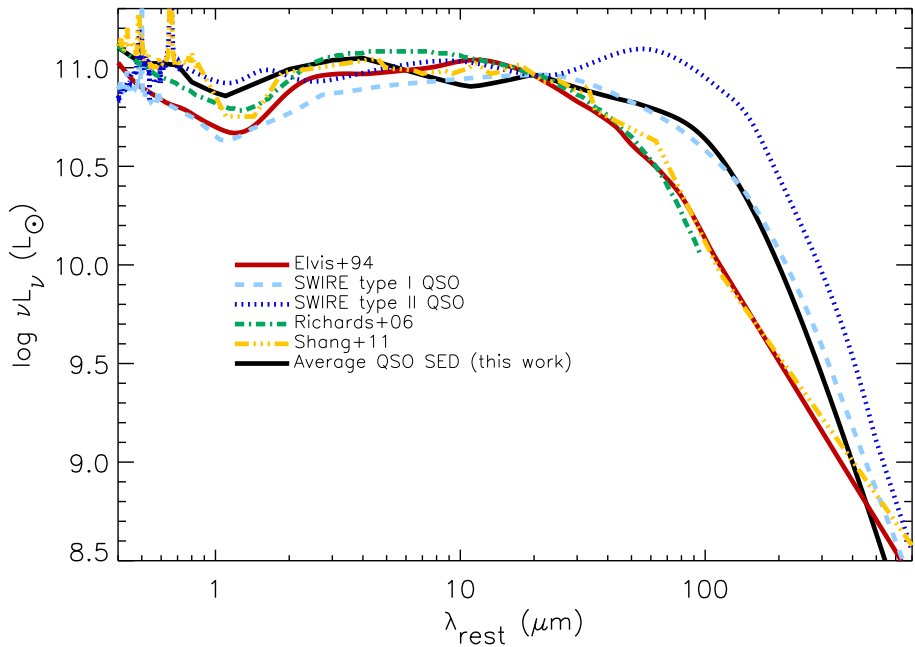,width=0.7\linewidth} 
\caption{The average QSO SED of our sample (black solid line), compared to those from Elvis et al. (1994), Richards et al. (2006),
 Shang et al. (2011) and the SWIRE template library are also included, normalised at 20$\mu$m.}
\label{fig:AGNtotalSED}
\end{figure*}

\begin{figure*}
\epsfig{file=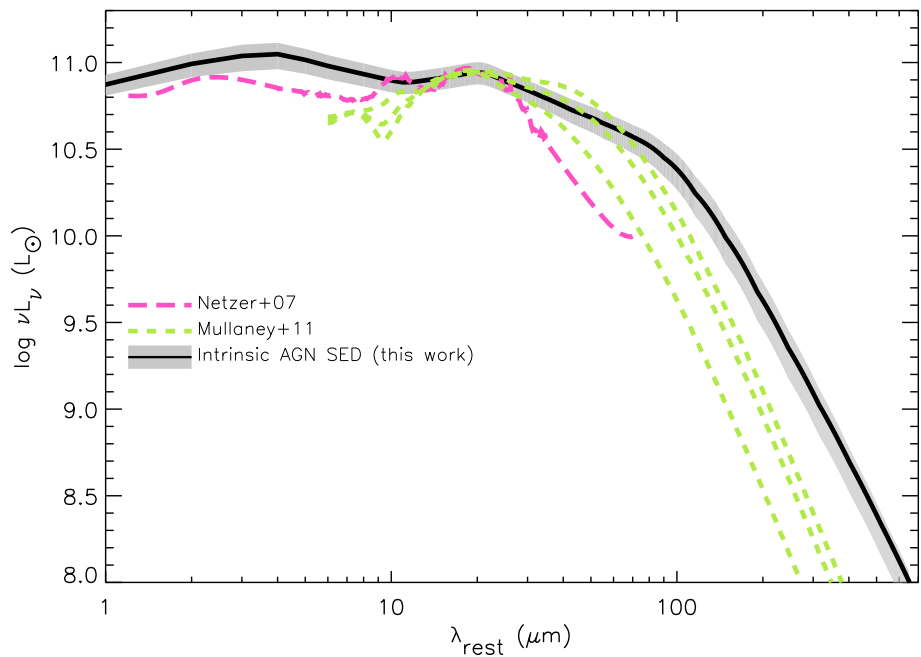,width=0.7\linewidth} 
\caption{Our intrinsic AGN SED (black solid line) and
 68 per cent confidence intervals (shaded region). For comparison the intrinsic AGN SEDs form Netzer et al. (2007) and Mullaney et al. (2011) are also shown, normalised at 20$\mu$m.}
\label{fig:AGNSED}
\end{figure*}

To summarise, a number of studies in the literature provide strong evidence that the 11.3\,$\mu$m PAH offers a robust measure of
star formation in AGN host galaxies, even in the circum-nuclear
regions. Only four of our sample of QSOs show the elevated
7.7\,$\mu$m/11.3\,$\mu$m PAH ratios which LaMassa
et~al. (2012\nocite{LaMassa12}) attribute to AGN-suppression of
11.3~$\mu$m PAH emission, and excising these objects from our sample
has no significant impact on our intrinsic AGN SED. There is no trend
in our sample between elevated 7.7\,$\mu$m/11.3\,$\mu$m PAH ratios and
AGN luminosity; indeed, most of the sources in our sample show
7.7\,$\mu$m/11.3\,$\mu$m PAH ratios which are consistent with those
found in star-forming galaxies. We also find no evidence for AGN
suppression of the 11.3\,$\mu$m emission when we compare the higher and
lower AGN-luminosity subsets of our sample. The star-formation
contribution to the average SED appears to be robust to any effect of
the AGN on the 11.3\,$\mu$m PAH emission, and hence the intrinsic AGN
SED appears to be likewise robust to any such effect.

\subsection{Comparison with other AGN SEDs}
\label{sec:otherAGNSEDs}

In Fig. \ref{fig:AGNtotalSED} we compare the average SED of our sample of QSOs (i.e. AGN+host, also shown in Fig. \ref{fig:SEDs_QSO_average}) to the average QSO SEDs from Elvis et
al. (1994\nocite{Elvis94}), Richards et al. (2006\nocite{Richards06}), Shang et al. (2011\nocite{Shang11}) and the type I QSO and type II QSO templates from the SWIRE
template library\footnote{http://www.iasf-milano.inaf.it/$\sim$polletta/templates/swire \textunderscore templates.html}, Polletta et al. 2006; 2007\nocite{Polletta06}\nocite{Polletta07}). 
Our QSO SED agrees well with the SWIRE type I QSO SED, particularly in the far-IR/submm, whereas it has less far-IR/submm power than the type II QSO SWIRE SED, likely due to a different amount of host galaxy contribution. On 
the other hand, the composite SEDs of Elvis et al. (1994), Richards et al. (2006) and Shang et al. (2011) fall short of our SED in the 60-300$\mu$m region. However, note that their data do not probe the submm and hence the SEDs 
are extrapolated in that wavelength range. Most of the Elvis et al. QSOs are not significantly detected at \textit{IRAS}/100$\mu$m and a large fraction ($\sim$25 per cent) are not significantly detected at \textit{IRAS}/60$\mu$m. 
Similarly, only a handful of the Richards et al. QSOs have 70$\mu$m data. With respect to the Shang et al. radio-quiet SED, again their data probe shorter rest-frame wavelengths both as a consequence of their sample extending to 
much higher redshift and the lack of far-IR photometry for about half of the sources.

Fig. \ref{fig:AGNSED} compares the intrinsic AGN SED extracted here to the SEDs  from Netzer et al. (2007\nocite{Netzer07}) and Mullaney et al. (2011\nocite{Mullaney11}), both of which are classed as pure AGN SEDs. The intrinsic 
AGN SED (in $\nu L_{\nu}$) we derive in this work can be considered `flat' within 0.2\,dex up to about 20$\mu$m, with a subsequent slow drop off of  $\sim$0.4\,dex from 20 to 60$\mu$m and $\sim$0.3\,dex from 60 to 100$\mu$m. 
Fig. \ref{fig:AGNSED} shows that the Netzer et al. and Mullaney et al. SEDs are much steeper in the far-IR than ours, particularly the former whose emission at 70$\mu$m is more than half a dex lower. The Netzer et al. AGN SED was 
derived for a subsample of PG QSOs which were studied as part of the \textit{Spitzer} QUASAR and ULIRG evolution study (QUEST; Schweitzer et al. 2006\nocite{Schweitzer06}). Netzer et al. produce a template starburst SED by 
nomalising the SEDs of 
the starburst-dominated ULIRGs in their sample at 60\,$\mu$m and then taking the mean at every wavelength. To determine the intrinsic AGN SED, they subtract the starburst template from the average PG 
QSO SED, first scaling it according to the assumption that most of the 50-100 $\mu$m emission from the PG QSOs is due to star formation. Note that our study suggests that this assumption is not valid. 
Mullaney et al. (2011) estimate the form of an intrinsic AGN SED by decomposing the observed SEDs of the sources in their sample into host-galaxy and intrinsic AGN components by simultaneously fitting the 
mid-IR spectra and \textit{IRAS} photometry. Their method however already assumes an intrinsic AGN component which is then used during spectral decomposition, created by combining the mid-IR spectra of AGN-dominated sources 
in the 6-25$\mu$m range with a modified blackbody above a wavelength $\lambda_{\rm BB}$, where $\lambda_{\rm BB}$ is allowed any value between 25\,$\mu$m and 100$\mu$m. The drawbacks of both Netzer et al. and Mullaney 
et al. methods is that they make a priori assumptions about the balance of AGN and stellar emission when calculating the intrinsic AGN SED. In contrast we argue that our method of extracting the pure AGN emission is more robust as 
it is not based in any inherent assumptions on the balance of AGN and stellar emission or the shape of the intrinsic AGN component.
The intrinsic AGN SED derived here maintains a higher level of far-IR emission compared to the Mullaney et al. and Netzer et al. SEDs. This far-IR excess likely stems from AGN heated dust in the host galaxy (at kpc scales), a component which is not accounted for in the Mullaney et al. and Netzer et al. SEDs.

\begin{figure}
\epsfig{file=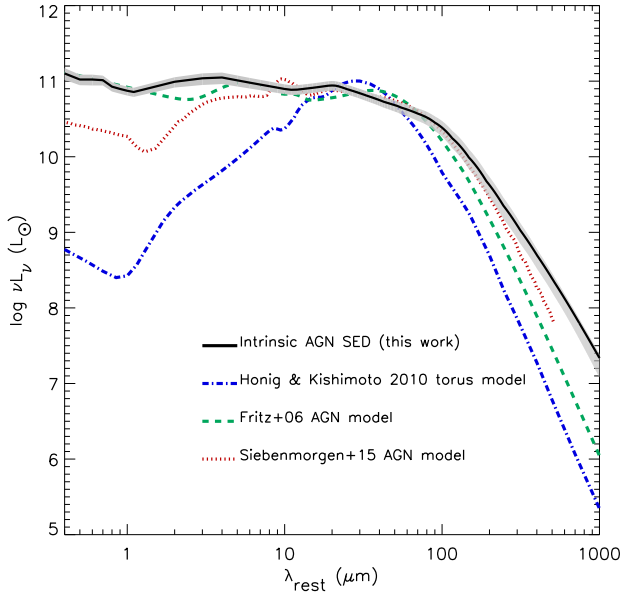,width=0.99\linewidth} 
\caption{Plot highlighting the differences between the intrinsic AGN
  SED (this work) and synthetic AGN models. The blue dot dashed line is the best-fit torus model from
  Honig $\&$ Kishimoto (2010), the green dashed line is the best fit
  model from Fritz et al. (2006) and the red dotted line is the best-fit model from Siebenmorgen et al. (2015).}
\label{fig:AGNtorus}
\end{figure}

\subsection{Comparison with AGN models}
In figure \ref{fig:AGNtorus} we compare our intrinsic AGN SED to torus models from Honig $\&$ Kishimoto (2010\nocite{HK10}, hereafter HK10) and accretion disk+torus models from Fritz et al. (2006\nocite{Fritz06}, hereafter F06) and 
Siebenmorgen et al. (2015\nocite{SHE15}, hereafter S15). Our aim is to examine whether the aforementioned widely used AGN models could reproduce our pure AGN SED in the far-IR/submm, resulting in the appropriate 
normalisation of a star-forming far-IR/submm component during multi-component SED fitting (e.g. see Hatziminaoglou et al. 2008 for multi-component SED fitting). 
The HK10 library of torus models is fit onto our pure AGN SED in the 3--100$\mu$m region, whereas the F06 and S15 libraries are fit from 0.4 to 100$\mu$m as they include the accretion disk component. In Fig. \ref{fig:AGNtorus}, we 
examine how well the best fit model (minimum $\chi^2$) from each library can reproduce the far-IR, $\lambda>$100\,$\mu$m, emission of our intrinsic AGN SED.

The best fit HK10 model has 0 inclination angle (type-I AGN), a 30
degree half opening angle of the torus, power-law index alpha of 0
(the lower alpha is the redder the SED), the highest number of clouds (N=10)
which gives maximum IR emission and the highest optical depth $\tau$=80
increasing the flux at longer wavelengths. From the F06 library the best fit model is for a face-on AGN, has the largest opening angle, no variation of the dust density in the vertical direction, a value of 10 for the equatorial optical depth
at 9.7$\mu$m and a ratio of 150 between the outer and inner radius of the dust distribution. The best-fit S15 model has an inner torus radius of 15.45$\times$10$^{17}$ cm (maximum in the library), a cloud volume filling factor of 1.5 
per cent, 0 optical depth in the individual clouds, maximum optical depth in the disk midplane (1000), and a viewing angle of 43 degrees measured from the z-axis.  

We see that the HK10 and F06 AGN models fail to reproduce the AGN power in the submm, whereas the S15 model does a better job, matching the mid to far-IR part of our SED up to about 200$\mu$m. This shortfall is likely due to the limited extent of the dust distribution in the formulation of these models and hence caution must be applied when using AGN models in SED multi-component fitting, as it could lead to overestimated SFRs.

\subsection{Implications for measuring star-formation rates in AGN host galaxies}
\label{sec:implications}

\begin{figure}
\epsfig{file=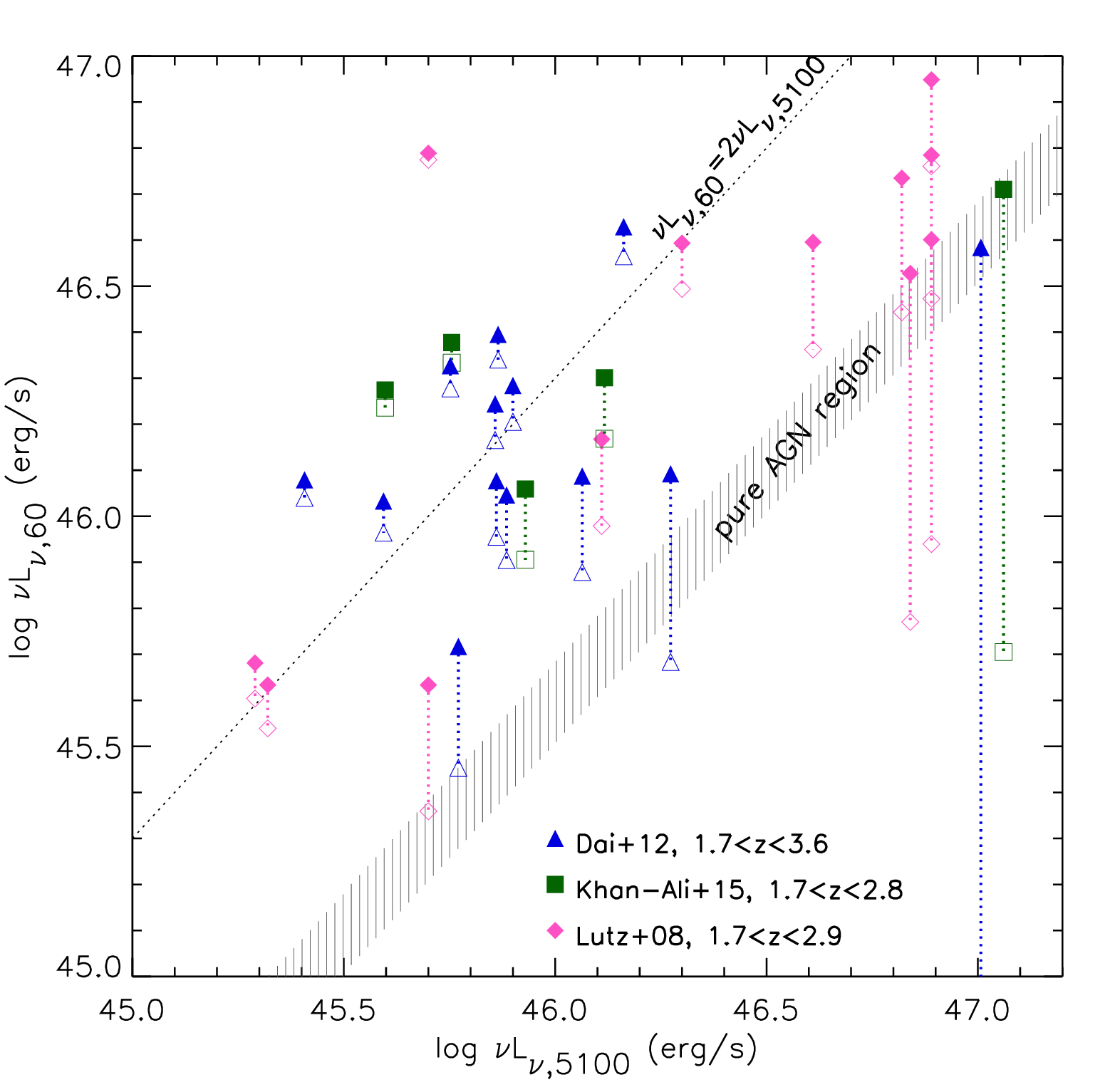,width=0.99\linewidth} 
\caption{60$\mu$m luminosity versus AGN power ($\nu L_{\nu, 5100}^{\rm AGN}$) for 3 luminous high redshift
QSO samples from the literature: Dai et al. (2012; $1.7<z<3.6$), Khan-Ali
et al. (2015; $1.7<z<2.8$) and Lutz et al. (2008; $1.7<z<2.9$). The grey
shaded region is the pure AGN line ($\pm$20 per cent uncertainty on the $\nu L_{\nu, 60}$/$\nu L_{\nu, 5100}$ ratio calculated by
bootstrapping). The filled-in symbols are the 60$\mu$m luminosity of each source ($\nu L_{\nu, 60}^{\rm tot}$)
whereas the open symbols are the 60$\mu$m luminosity attributed to star-formation with the AGN contribution subtracted ($\nu L_{\nu, 60}^{\rm SF}$). The dotted line is at $\nu L_{\rm \nu, 60}^{\rm tot}$/$\nu L_{\nu, 5100}^{\rm 
AGN}$$=$2 and represents the ratio above which the AGN contribution to the far-IR SED may be ignored.}
\label{fig:Rosario12_plot_wothersamples}
\end{figure}

Our work suggests that for AGN hosts, particularly those of powerful AGN, the contribution from the AGN to the infrared/submm needs to be taken into account when calculating properties such as SFRs. In Fig. 
\ref{fig:Rosario12_plot_wothersamples} we use 3 luminous QSO samples from the literature to demonstrate the level of correction that would be required when measuring the star-forming emission in luminous AGN. We use type 1 
radio-quiet QSOs with robust submm/mm detections at $1.7<z<2.9$ from Lutz et al. (2008)\nocite{Lutz08}, 24$\mu$m-selected broad-line QSOs at $1.7<z<3.6$ from Dai et al. (2012)\nocite{Dai12} and X-ray absorbed and submm-
luminous type I QSOs at $1.7<z<2.8$ from Khan-Ali et al. (2015\nocite{Khan-Ali15}). The 60$\mu$m luminosity of each source ($\nu L_{\rm \nu, 60}^{\rm tot}$) is shown as a function of $\nu L_{\nu, 5100}^{\rm AGN}$, using their 
published values or values obtained by private communication. The shaded region is our pure AGN SED, its width corresponding to the 20 per cent error on the $\nu L_{\nu, 60}$/$\nu L_{\nu, 5100}$ ratio calculated by bootstrapping. 
Using our pure AGN SED, we subtract the expected AGN contribution to $\nu L_{\rm \nu, 60}^{\rm tot}$, obtaining the luminosity at 60$\mu$m which can solely be attributed to star-formation ($\nu L_{\rm \nu, 60}^{\rm SF}$). Note that 
in many cases $\nu L_{\rm \nu, 60}^{\rm SF}$ is up to an order of magnitude lower than $\nu L_{\rm \nu, 60}^{\rm tot}$. Fig. \ref{fig:Rosario12_plot_wothersamples} illustrates that one could assume a negligible AGN contribution only if 
the total galaxy emission at 60$\mu$m is more than a factor of two higher than the intrinsic AGN power measured in the optical, i.e. $\nu L_{\rm \nu, 60}^{\rm tot}$/$\nu L_{\nu, 5100}^{\rm AGN}$$>$2. In fact, the AGN 
contribution cannot be simply ignored even at $\lambda>200$\,$\mu$m; the third panel of Fig. \ref{fig:lumQSO_SED_average_wDH02} shows that in our sample, the average AGN SED is at least comparably luminous to the average 
SF SED at all wavelengths suggesting that, in some galaxies hosting luminous AGN, there is no `safe' broadband photometric observation ($\lesssim1000\mu$m) that can be assumed free from AGN contamination. Assuming an M82 
SED (from the SWIRE SED library) where $L_{\rm IR, 8-1000\mu m}$=1.7$\nu L_{\nu, 60}$, broadly speaking, the AGN contribution could be ignored if the total IR emission from a given galaxy is more than a factor of 4 higher than the intrinsic AGN 
power at 5100\AA.

A further point to note is that submm colours are not necessarily good indicators of whether the galaxy is AGN or SF dominated in the far-IR/submm. Although the pure AGN SED and the SEDs of star-forming galaxies are distinctly 
different at $\lesssim$100\,$\mu$m, they decay with a similar rate in the submm (e.g. see Fig. \ref{fig:lumQSO_SED_average_wDH02}), as in both cases we are into the Rayleigh-Jeans part of the modified blackbody emission from 
dust in the host galaxy. As a result we would expect that all SEDs would have consistent submm colours. Indeed we find that the rest-frame \textit{Herschel}/SPIRE 500/350$\mu$m and 350/250$\mu$m colours of our pure AGN SED 
are within the range probed by nearby spiral galaxies (Boselli et al. 2010\nocite{Boselli10}). Redshifting the AGN SED to $z=2$, the observed SPIRE colours now coincide with the peak in the observed distribution of colours in SPIRE-
selected galaxies and AGN as shown in Hatziminaoglou et al. (2010). This implies that the rest-frame submm colours (or observed submm colours at low redshift ) of AGN hosts are not an indication of whether AGN or star-formation is 
the dominant energy source in the submm.

\begin{figure}
\epsfig{file=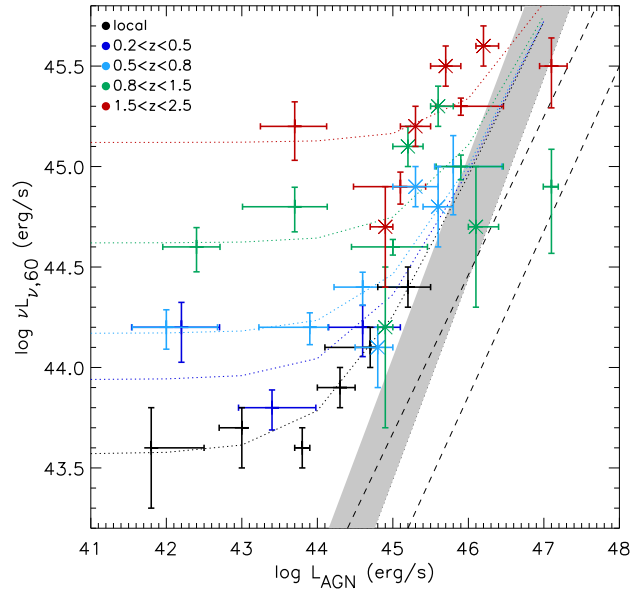,width=0.99\linewidth} 
\caption{Average 60$\mu$m rest-frame luminosity versus AGN luminosity from Rosario et al. (2012; crosses and dotted lines) and (2013; asterisks). The
  colours represent different redshift bins as shown in the
  legend. The region bound by the dashed lines is the Rosario et al. estimate of a pure AGN contribution. The shaded
  region is the pure AGN region derived in this work, its width incorporating the range in $\nu L_{\nu, 5100}$ to $L_{\rm AGN}$
conversion factors (see section \ref{sec:implications}) together with the 20 per cent uncertainty in $\nu L_{\nu, 60}$/$L_{\rm AGN}$ that we calculate from bootstrapping.}
\label{fig:Rosarioplot}
\end{figure}

We next examine the implications of our pure AGN SED for the statistical studies of star-formation in X-ray selected AGN from Rosario et al (2012; 2013) and Harrison et al. (2012). 
Rosario et al. (2012\nocite{Rosario12}; 2013\nocite{Rosario13}) examine the star-forming properties of X-ray selected AGN hosts, using the 60$\mu$m luminosity as a proxy for star-formation and assuming a negligible 
contribution from the AGN at 60$\mu$m; in Fig. \ref{fig:Rosarioplot} we show the average luminosities for the Rosario et al. samples as well as the functional fits to them. Also plotted is the region in parameter 
space that corresponds to the intrinsic AGN emission in their work, derived by using the AGN SEDs of Mullaney et al. (2011) and Netzer et al. (2007). The shaded region represents the $L_{\rm 60}$/$L_{\rm 
AGN}$ ratio that stems from our intrisic AGN
SED.  As in Rosario et al. (2012; 2013), to convert from $\nu L_{\nu, 5100}$ to $L_{\rm AGN}$, where $L_{\rm AGN}$ is the bolometric AGN luminosity, 
we use the Netzer and Trakhtenbrot (2007\nocite{NT07}) relation 
$L_{\rm AGN}$=$f$$\times$$\nu L_{\nu, 5100}$, where $f=7$. However as Netzer $\&$
Trakhtenbrot (2007\nocite{NT07}) point out, estimates for $f$
in type 1 AGN are in the range 5-13 (e.g., Elvis et al. 1994; Kaspi et al. 2000\nocite{Kaspi00}; Netzer 
2003\nocite{Netzer03}; Marconi et al. 2004\nocite{Marconi04}). The range in $\nu L_{\nu, 5100}$ to $L_{\rm AGN}$
conversion factors quoted above are incorporated into the width of the intrinsic AGN region together with the 20 per cent uncertainty in
$\nu L_{\nu, 60}$/$L_{\rm AGN}$ that we calculate from
bootstrapping.

\begin{figure}
\epsfig{file=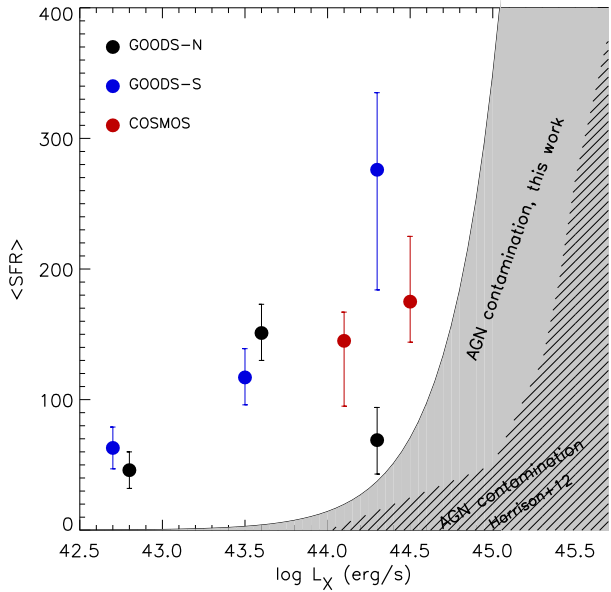,width=0.99\linewidth} 
\caption{Average SFR versus X-ray AGN luminosity for the X-ray
  selected sample of AGN at z=1-3 in Harrison et al. (2012). The regions of AGN contamination are also shown - shaded area: this work, hatched area: Harrison et al. (2012)}
\label{fig:Harrisonplot}
\end{figure}

Note that our pure AGN SED shifts the $\nu L_{\nu, 60}$/$L_{\rm AGN}$ ratio to higher values, i.e. the AGN contribution at 60$\mu$m is now 2--10 times higher than what is assumed in the work of Rosario et al. (2012; 2013). Rosario et al. interpret 
their results as two different mechanisms at play regarding the black hole/SF connection, originally proposed by Shao et al. (2010): they suggest that
for low AGN luminosities the SFR does not correlate with AGN power, so secular mechanisms drive the black hole and galaxy growth, whereas at
high AGN luminosities there is a correlation, a sign of the increasing importance of major-mergers in driving both the growth of super-massive black holes and star-formation. 
Looking at Fig. \ref{fig:Rosarioplot} and the location of the pure AGN region derived in this work, we arrive at a different interpretation of the Rosario et al. results: we propose that the difference in $\nu L_{\nu, 60}/L_{\rm AGN}$ 
between the low and high
luminosity AGN can be understood simply as an increase in the AGN contribution. In Fig. \ref{fig:Rosarioplot} it is evident that one can only travel so far in
the x-direction before the pure AGN region (grey shaded region) is met, so for a given level of $\nu L_{\nu, 60}$, low luminosity AGN can have a higher fraction of
star-formation contribution than their higher luminosity
counterparts. In other words, at any given redshift, and hence mean SFR, the more luminous AGN will on average have higher
$\nu L_{\nu, 60}$ simply due to increased AGN emission at 60$\mu$m. Our results indicate that the apparent correlation between $\nu L_{\nu, 60}$ and $L_{\rm AGN}$ in high luminosity AGN is driven by the intrinsic AGN emission at those wavebands, also corroborated by the convergence of the $\nu L_{\nu, 60}$--$L_{\rm AGN}$ trends (dotted lines) onto the intrinsic AGN region (grey shaded region) at high $L_{\rm AGN}$.

Figure \ref{fig:Harrisonplot} shows the results from Harrison et
al. (2012\nocite{Harrison12}), who examined the average SFR in a sample of luminous X-ray
selected AGN at $z=1-3$. The SFRs presented are not corrected for AGN contamination and the parameter space assigned to the AGN contribution is shown. Following their method, we also calculate a region 
of AGN contamination, first using our pure AGN SED to obtain the rest-frame luminosity
at 83$\mu$m (250$\mu$m observed at $z=2$) and
then converting it to an SFR using the Elbaz et al. (2011\nocite{Elbaz11}) main sequence
template and the Kenniccutt (1998\nocite{Kennicutt98}) relation. To convert from $\nu L_{\nu, 5100}$ to $L_{\rm X}$ we use the Maiolino et al. (2007) relation, log\,$L_{\rm X}(\rm 2-10keV)=0.721\times \rm log (\nu L_{\nu, 5100})
+11.78$, 
where the units are in erg/s. The region of AGN contamination derived using our pure AGN SED extends further than the one shown in Harrison et al., implying a correction in the star-formation rates would be 
necessary for the most luminous AGN. Indeed correcting for AGN contamination might bring these results in agreement with Page et al. (2012) who find that the most luminous AGN do not lie in the most highly star-forming hosts (see also Barger et al. 2015).

\section{Summary $\&$ Conclusions}
\label{sec:conclusions}
We have explored the question of whether and how much the AGN contributes to the far-IR emission from a galaxy. For our purposes we used a sample of 47 broad line, luminous ($L_{\rm 5100}>$10$^{43.5}$\, 
erg/s), $z<0.18$, radio-quiet QSOs from the Palomar Green survey. Taking advantage of archival data in the 0.4-500$\mu$m range and PAH-derived estimates of star-forming luminosities, we removed the 
stellar component in these sources' energy budget using the 11.3$\mu$m PAH feature in the QSOs' mid-IR spectra, in order to obtain an average \emph{intrinsic} or \emph{pure} AGN SED in the optical--submm wavelength range. The 1$\sigma$ uncertainty on the intrinsic AGN SED ranges between 12 and 45 per cent as a function of wavelength and is a combination of PAH flux measurement errors and the uncertainties related to the conversion between PAH luminosity and star-forming luminosity.

Our main conclusions are:
\begin{itemize}

\item  In our sample of QSOs, the average AGN contribution to the total energy budget is $>95$ per cent shortwards of 20$\mu$m, $60-95$ per cent in the 20--100$\mu$m range and at all wavelengths thereafter (at least up to 1000$\mu$m) it is comparable to the contribution from star-formation. This implies that for some galaxies hosting powerful AGN, the entire optical-submm broadband SED can be AGN dominated and hence there is no `safe' waveband (at $\lambda <1000 \mu$m) that can be used in calculating SFRs without subtracting the AGN contribution. 

\item Our intrinsic AGN SED maintains a higher level of far-IR emission compared to other such SEDs available in the literature. This far-IR excess likely stems from AGN heated dust in the host galaxy (at kpc scales), a component which had been previously ignored. 

\item Using AGN torus models in AGN/SF SED decomposition could potentially underestimate the far-IR contribution of the AGN and lead to overestimated SFRs, as many of these models are built with a limited extent in the dust distribution. 

\item The shape of the far-IR component of the intrinsic AGN SED does not vary with AGN luminosity and our AGN template can be considered applicable to any samples of galaxies hosting a luminous ($L_{\rm 5100}$ or $L_{\rm X(2-10keV)}$ $\gtrsim$10$^{43.5}$ erg/s) AGN.

\item Statistical studies of star-formation in AGN host galaxies need to incorporate a correction for the AGN contribution, irrespective of the waveband used to calculate SFRs. This is particularly crucial for 
luminous ($L_{\rm 5100}$ or $L_{\rm X(2-10keV)}$ $\gtrsim$10$^{43.5}$ erg/s) AGN. For ease, and broadly speaking, the AGN contamination could be neglected if the intrinsic AGN power at 5100$\rm \AA$ is more than a factor of 2 lower than the galaxy's 60$\mu$m luminosity and more than factor of 4 lower than the total IR emission (8--1000$\mu$m) of the galaxy.  
\end{itemize}

\clearpage
\appendix

{\bf{APPENDIX A: QSO SEDs}}
\label{appendixA}

\begin{figure*}
\epsfig{file=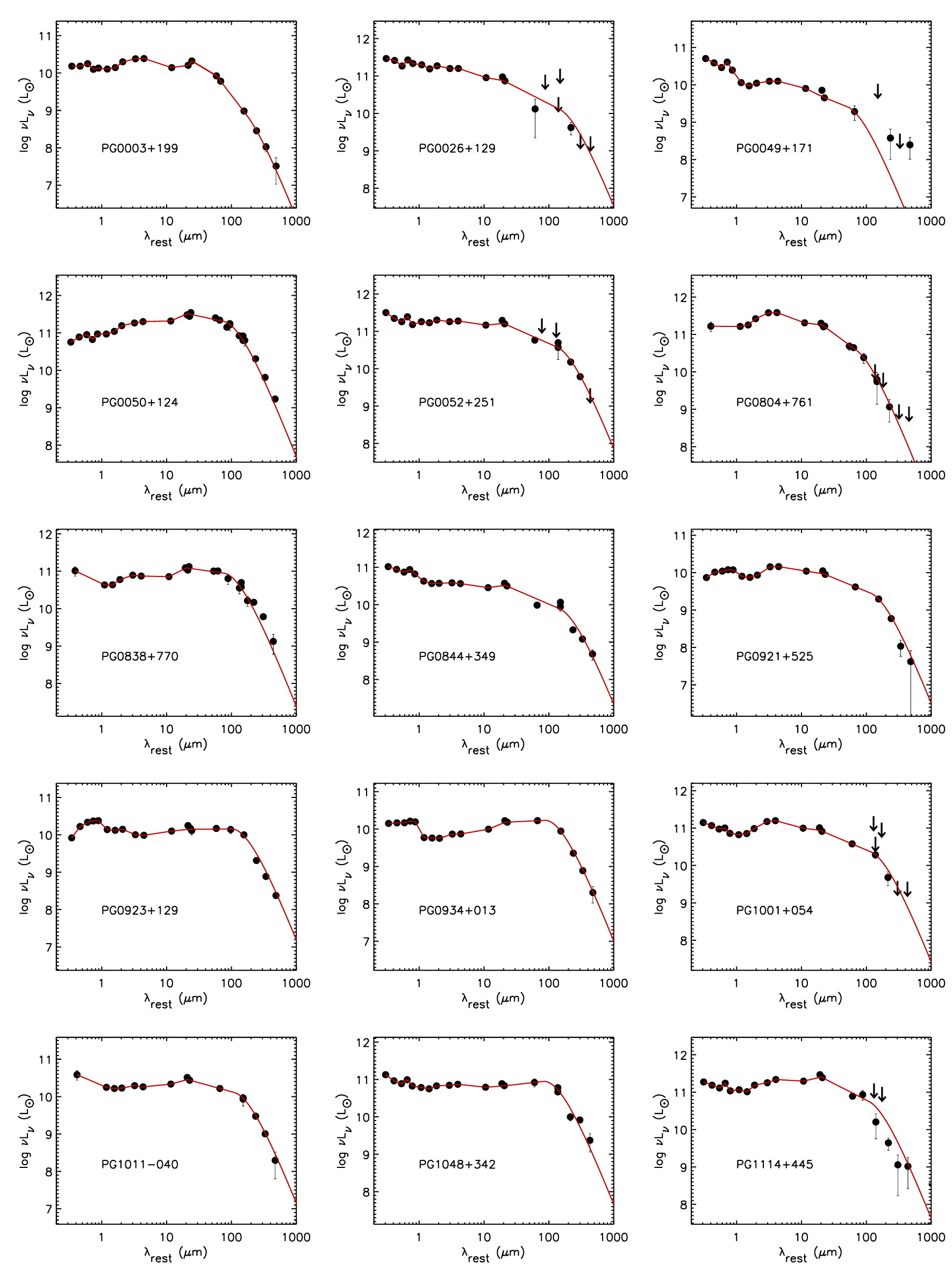,width=0.97\linewidth} 
\caption{The rest-frame SED of each QSO in our final sample. Photometry is denoted with black filled in circles if $>3\sigma$ and downward arrows if upper limits, and our fit to the data (see section \ref{sec:SED_fitting}) is shown as a 
red curve.}
\label{fig:seds1}
\end{figure*}

\begin{figure*}
\epsfig{file=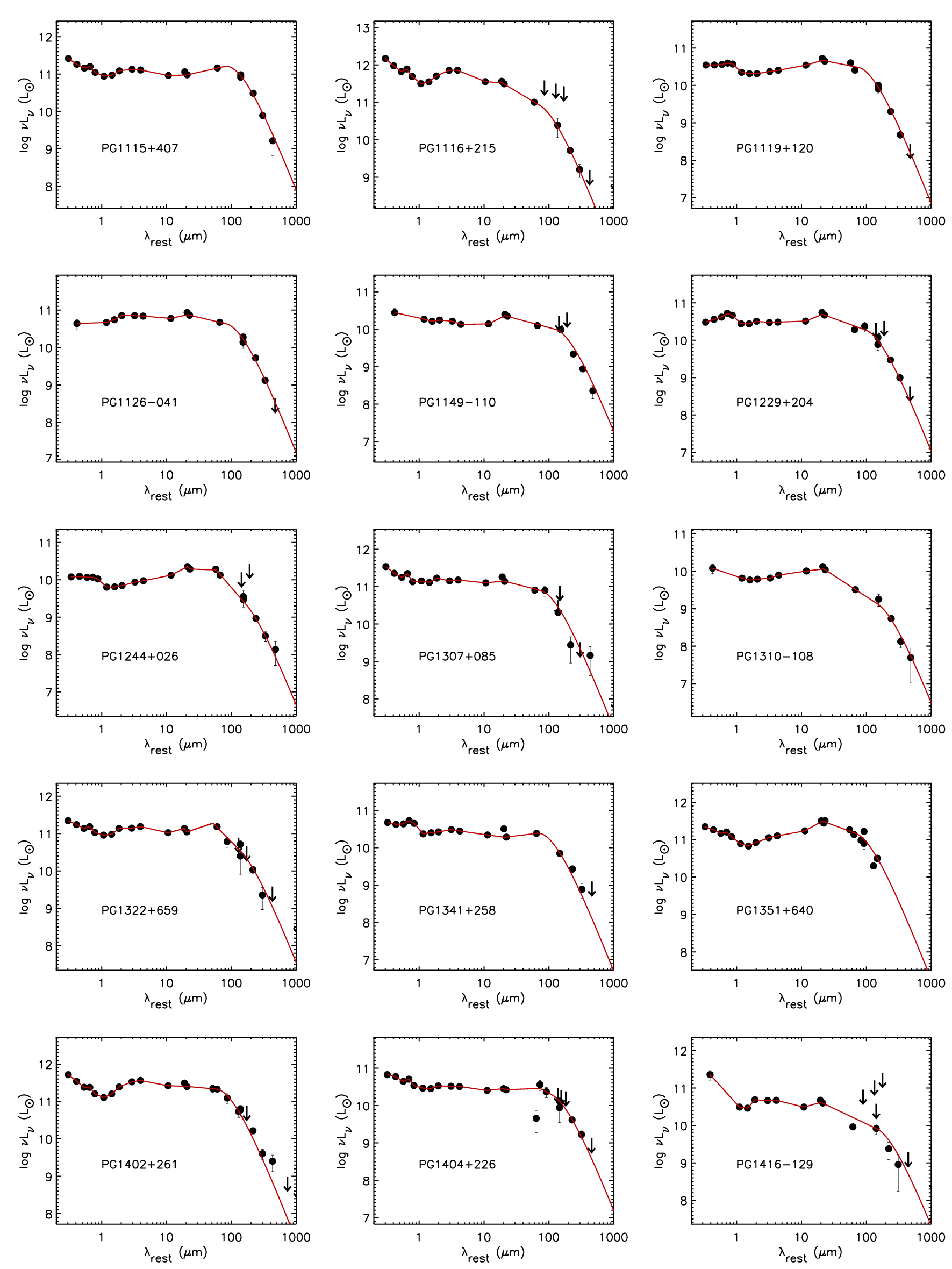,width=0.97\linewidth} 
\caption{The rest-frame SED of each QSO in our final sample. Photometry is denoted with black filled in circles if $>3\sigma$ and downward arrows if upper limits, and our fit to the data (see section \ref{sec:SED_fitting}) is shown as a 
red curve.}
\label{fig:seds2}
\end{figure*}

\begin{figure*}
\epsfig{file=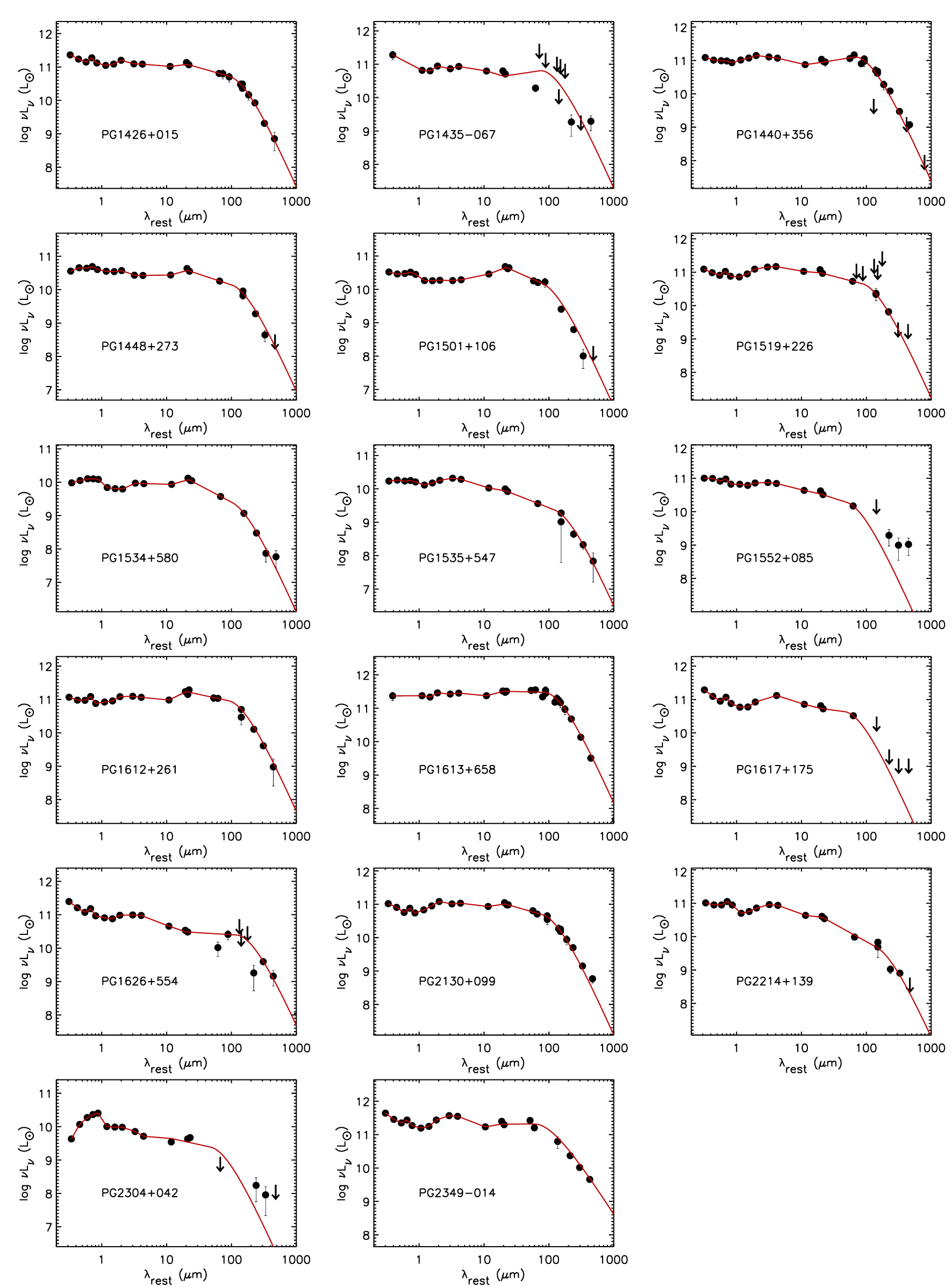,width=0.97\linewidth} 
\caption{The rest-frame SED of each QSO in our final sample. Photometry is denoted with black filled in circles if $>3\sigma$ and downward arrows if upper limits, and our fit to the data (see section \ref{sec:SED_fitting}) is shown as a 
red curve.}
\label{fig:seds3}
\end{figure*}

\clearpage
{\bf{APPENDIX B: SPIRE DATA}}
\label{appendixB}
\begin{table}
\centering
\caption{SPIRE flux densities and corresponding 1$\sigma$ uncertainty (in mJy) }
\begin{tabular}{lc|c|c|c|c|c|c|c|}
\hline 
name&$f_{250}$&$\sigma_{250}$&$f_{350}$&$\sigma_{350}$&$f_{500}$&$\sigma_{500}$\\
PG0003+199&64.01&9.77&33.14&8.57&14.52&9.71\\
PG0007+106&160.55&8.49&168.07&7.77&150.99&9.87\\
PG0026+129&24.77&8.90&0.00&7.72&0.00&9.56\\
PG0049+171&12.16&8.89&0.00&7.42&16.01&9.36\\
PG0050+124&724.67&11.80&322.41&8.65&121.57&9.58\\
PG0052+251&73.89&8.92&41.93&7.40&2.55&8.91\\
PG0157+001&497.43&12.46&180.65&8.22&66.38&9.82\\
PG0804+761&14.61&8.86&1.54&7.45&0.00&9.87\\
PG0838+770&103.76&9.00&59.91&7.84&18.70&10.19\\
PG0844+349&68.59&9.64&54.56&7.68&30.84&9.61\\
PG0921+525&66.49&8.39&17.02&7.83&9.41&9.29\\
PG0923+201&3.09&8.76&0.65&7.81&14.94&9.41\\
PG0923+129&340.31&9.52&178.35&8.25&78.63&9.55\\
PG0934+013&121.91&9.08&58.27&7.55&21.64&10.18\\
PG1001+054&21.85&8.77&1.92&7.65&5.33&9.35\\
PG1011-040&118.77&8.57&56.09&7.53&15.61&10.57\\
PG1012+008&49.64&8.74&18.32&7.58&0.00&9.96\\
PG1022+519&127.28&9.66&46.67&7.51&32.21&10.14\\
PG1048+342&41.24&9.01&47.84&7.49&19.49&9.93\\
PG1114+445&25.69&9.26&9.28&7.90&12.12&9.01\\
PG1115+407&152.57&8.91&53.99&7.23&16.32&9.68\\
PG1116+215&19.09&3.69&8.27&3.14&0.00&3.78\\
PG1119+120&108.01&8.75&36.23&7.63&1.13&9.48\\
PG1126-041&195.00&9.31&68.28&7.56&4.05&9.65\\
PG1149-110&122.67&9.72&68.75&8.06&25.44&9.82\\
PG1151+117&9.25&9.11&0.14&8.24&0.00&10.02\\
PG1202+281&36.75&9.08&18.15&7.43&15.60&9.47\\
PG1211+143&11.46&9.05&5.57&7.22&18.79&9.79\\
PG1226+023&455.31&6.52&692.08&7.78&1025.31&10.52\\
PG1229+204&99.07&9.22&46.38&7.77&9.32&9.87\\
PG1244+026&54.49&8.38&25.85&7.72&16.14&10.12\\
PG1307+085&13.40&8.96&0.21&7.40&14.14&10.09\\
PG1309+355&76.71&9.43&37.04&7.30&40.05&9.89\\
PG1310-108&65.23&8.61&22.21&7.26&11.75&9.27\\
PG1322+659&43.94&8.52&12.99&7.66&0.00&10.40\\
PG1341+258&45.67&8.93&18.23&7.83&10.20&10.72\\
PG1351+236&208.28&9.30&75.56&7.53&13.74&9.29\\
PG1351+640&0.00&0.00&0.00&0.00&0.00&0.00\\
PG1352+183&8.73&9.29&1.91&7.66&0.84&10.18\\
PG1402+261&70.79&8.46&24.21&7.59&21.69&10.18\\
PG1404+226&54.39&8.74&30.88&7.79&3.99&10.37\\
PG1415+451&67.04&8.48&27.69&7.90&0.00&9.80\\
PG1416-129&17.30&8.39&9.20&7.42&0.00&9.28\\
PG1426+015&146.11&12.13&49.94&9.70&24.64&13.89\\
PG1435-067&14.25&8.94&4.02&9.13&29.97&14.35\\
PG1440+356&251.84&8.67&86.19&7.49&49.24&9.27\\
PG1448+273&59.39&9.75&19.30&7.31&1.81&8.92\\
PG1501+106&66.20&8.69&15.08&8.74&1.86&13.69\\
PG1519+226&42.05&9.33&4.48&7.87&0.00&11.99\\
PG1534+580&49.69&8.14&17.17&7.72&19.34&10.31\\
PG1535+547&42.04&8.90&28.18&7.41&12.99&9.94\\
PG1552+085&16.97&8.84&12.04&7.75&18.23&9.84\\
PG1612+261&91.51&8.90&40.95&7.36&13.59&9.92\\
PG1613+658&346.13&10.36&137.77&8.00&46.72&9.07\\
PG1617+175&5.97&8.54&0.00&7.84&3.32&10.09\\
PG1626+554&12.32&8.71&37.58&7.32&19.76&9.71\\
PG2130+099&173.01&8.51&68.23&7.28&40.54&10.24\\
PG2209+184&117.03&9.06&81.58&7.57&74.22&10.09\\
PG2214+139&33.01&8.25&35.28&7.40&8.41&9.51\\
PG2304+042&13.51&9.16&9.81&7.46&0.00&9.36\\
PG2349-014&88.42&3.75&54.83&3.09&34.49&3.93\\
\hline
\end{tabular}
\label{table:spire_fluxes}
\end{table}

\clearpage

\noindent
{\bf{APPENDIX C: TESTING THE METHOD OF SFR ESTIMATION ON STARBURST GALAXIES}}
\label{appendixC}
\vspace{3mm}

\begin{figure}
\epsfig{file=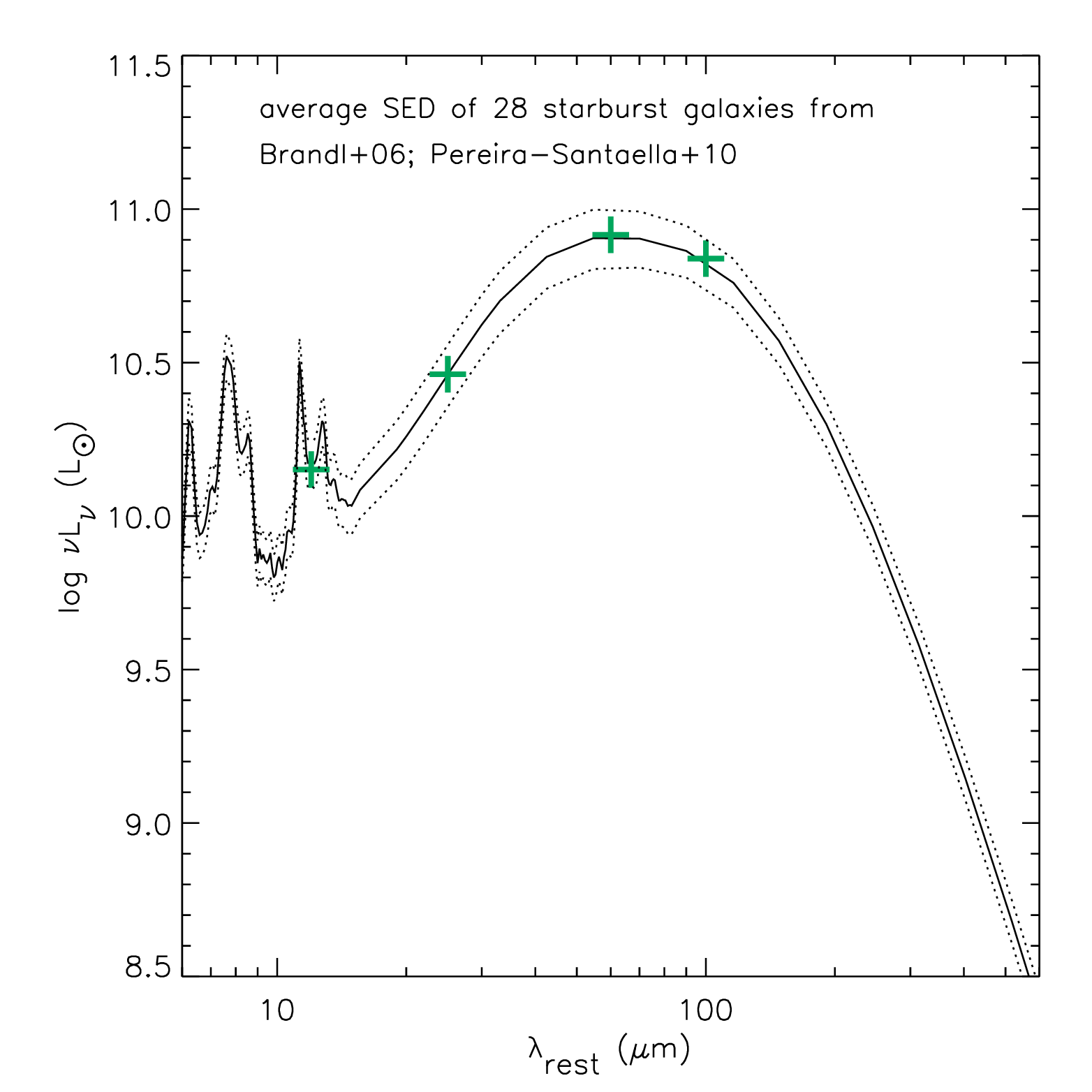,width=0.99\linewidth} 
\caption{The average template SED of sample A (28 sources; Brandl et al. 2006, Pereira-Santaella et al. 2010; solid curve and dotted curves for the 1$\sigma$ uncertainty calculated by bootstrapping) compared with the true average luminosities of this sample from \textit{IRAS} photometry  at 12, 25, 60 and 
100$\mu$m (green crosses). There is excellent agreement between the true \textit{IRAS}-derived average SED and the average SED derived using our method from the PAH measurements.}
\label{fig:test1}
\end{figure}

\begin{figure}
\epsfig{file=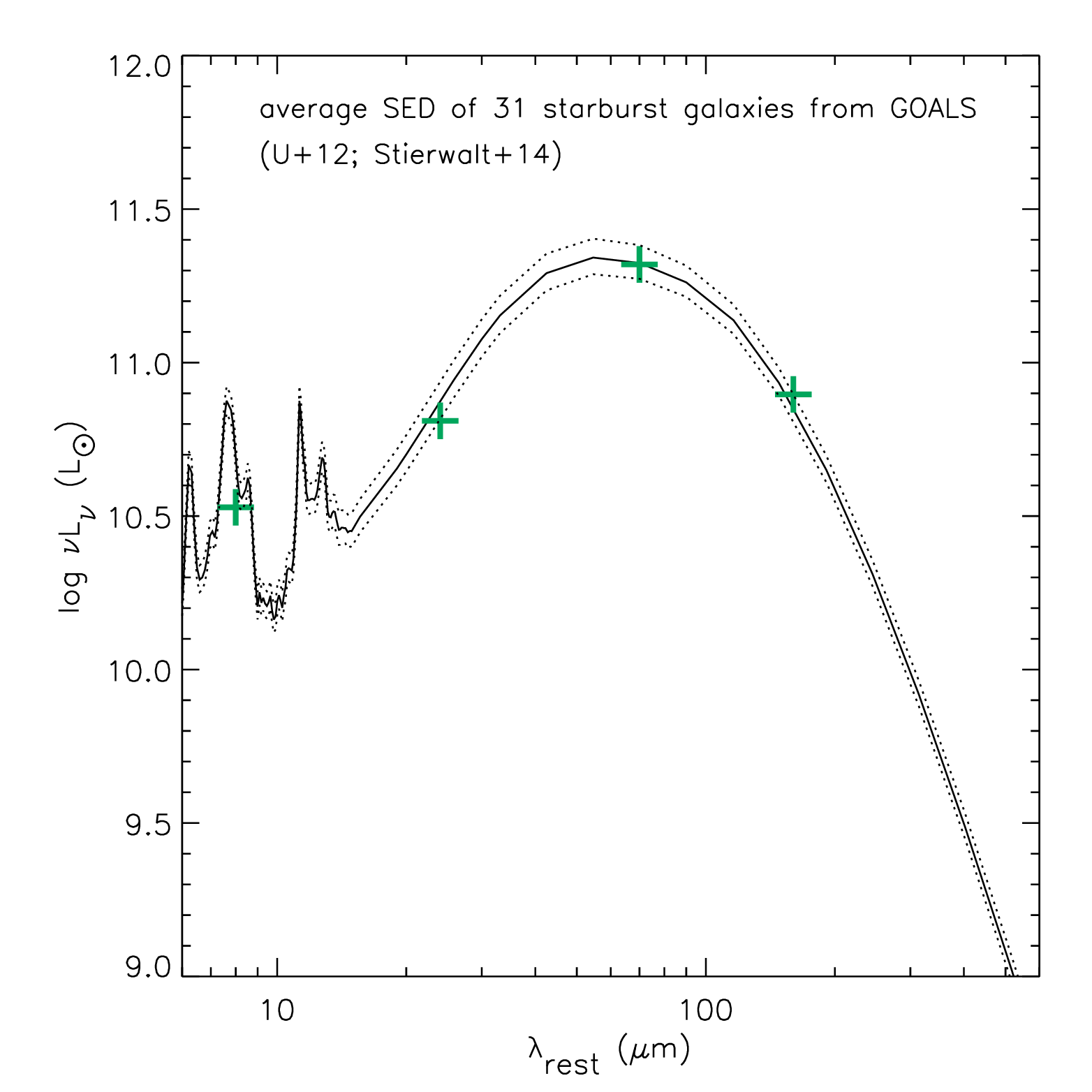,width=0.99\linewidth} 
\caption{The average template SED of sample B (31 sources; U et al. 2012 and Stierwalt al. 2014; solid curve and dotted curves for the 1$\sigma$ uncertainty calculated by bootstrapping), compared with the true average luminosities of this sample from \textit{Spitzer} photometry  at 8, 24, 70 and 160$\mu$m 
(green crosses). There is excellent agreement between the true \textit{Spitzer}-derived average SED and the average SED derived using our method from the PAH measurements.}
\label{fig:test2}
\end{figure}

\noindent As explained in section \ref{sec:selection_criteria}, we take the value of $L_{\rm SFIR}$ for each QSO from Shi et al. (2007). Shi et al. measure the strength of the 7.7 and 11.3$\mu$m PAHs in the QSO spectra and derive the 
conversion factor between PAH flux and 8-1000$\mu$m total infrared luminosity attributed to star-formation ($L_{\rm SFIR}$) for each PG QSO, by adopting an SED template from the Dale $\&$ Helou (2002\nocite{DH02}) library that 
gives the closest PAH flux at the redshift of the object. Here we test this method on some samples of nearby staburst galaxies using the 11.3$\mu$m PAH, which is the feature used in deriving $L_{\rm SFIR}$ for the vast majority of our 
sample. 
\begin{enumerate}
\item Following the method of Shi et al. (2007), we assign total infrared luminosities ($L_{\rm IR}$; 8--1000$\mu$m) to the Dale $\&$ Helou (2002; DH02) SED library using the 60/100$\mu$m colour and the relation in Marcillac et al. 
(2006), log\,($f_{60}/f_{100}$)=0.128\,log\,$L_{\rm IR}$-1.611. Each SED template is subsequently normalised to units of L$_{\odot}$ by its $L_{\rm IR}$.
\item We measure the luminosity of the 11.3 PAH (\textit{single feature}) in each DH02 template by integrating under the feature in the 10.9--11.8$\mu$m region and subtracting the continuum in the same region (the continuum 
baseline estimated by interpolating between 10.9--11.8$\mu$m). We also measure the luminosity of the PAH 11.3 \textit{complex} (as in Stierwalt et al. 2014) by integrating in the 10.9--13.1$\mu$m wavelength range. 
\item For the first comparison we use starburst galaxies from Brandl et al. (2006) and Pereira-Santaella et al. (2010), 28 sources in total (we refer to this as sample A). We take the 11.3$\mu$m single feature PAH luminosities published 
in these papers and \textit{IRAS} far-IR fluxes from NED, the latter simply converted to rest-frame luminosities, with no $K$-corrections required as these are nearby galaxies. For the second comparison we use 31 starburst galaxies 
from GOALS (we refer to this as sample B) which have published \textit{Spitzer}/MIPS fluxes in U et al. (2012) and 11.3 complex PAH fluxes in Stierwalt et al. (2014). Again these are nearby galaxies so no $K$-corrections are 
necessary. Note that all galaxies are larger than the \textit{Spitzer}/IRS slit, so we correct the published PAH luminosities according to the factors tabulated in these papers.
\item Each starburst galaxy is then assigned the DH02 template which has the closest PAH luminosity to the true PAH luminosity of the source  (single feature for sample A, complex feature for sample B). We then average the 
templates matched to sample A and B separately and compare with the average true far-IR SEDs built using the \textit{IRAS} and \textit{Spitzer} photometry. The 1$\sigma$ uncertainties are calculated by boostrapping the matched 
SED templates for each sample. The results are shown in Fig. \ref{fig:test1} and \ref{fig:test2}. There is excellent agreement between the true \textit{Spitzer} or \textit{IRAS}-derived SEDs and the average SEDs derived using our method from the PAH measurements.

\end{enumerate}

\clearpage

{\bf{APPENDIX E: The intrinsic AGN SED}}
\label{appendixE}
\vspace{3mm}

\begin{table}
\centering
\caption{Intrinsic AGN SED including the 68 per cent confidence upper and lower bounds. This table can be found in the online version of the journal, only the first few entries are shown here.}
\begin{tabular}{lc|c|c|c|}
\hline 
$\lambda_{\rm rest}$&log\,[$\nu L_{\nu}$]& log\,[$\nu L_{\nu, upper}]$& log\,[$\nu L_{\nu, lower}]$\\
 $\mu$m&L$_{\odot}$ & L$_{\odot}$& L$_{\odot}$\\
\hline
     0.4& 11.10 & 11.17 & 11.02\\
      0.5& 11.02 & 11.09 & 10.95\\
      0.6& 11.02 & 11.09 & 10.94\\
      0.7& 11.01 & 11.07 & 10.94\\
      0.8 & 10.93 & 10.98 & 10.86\\
      0.9& 10.90 & 10.95 & 10.83\\
       1 & 10.87 &  10.93 &  10.81\\
       2 & 10.99 &  11.05 &  10.92\\
       3 & 11.04 &  11.10 &  10.95\\
       4 & 11.05 &  11.11 &  10.96\\
       5 & 11.02 &  11.08 &  10.93\\
       6 & 10.98 &  11.04 &  10.90\\
       7 & 10.96 &  11.02 &  10.88\\
       8 & 10.93 &  11.00 &  10.86\\
       9 & 10.91 &  10.98 &  10.84\\
      10 & 10.90 &  10.96 &  10.83\\
 \hline
\end{tabular}
\label{table:residual_sed}
\end{table}

\section*{Acknowledgments}
The Herschel spacecraft was designed, built, tested, and launched under a contract to ESA managed by the Herschel/Planck Project team by an industrial consortium under the overall responsibility of the prime 
contractor Thales Alenia Space (Cannes), and including Astrium (Friedrichshafen) responsible for the payload module and for system testing at spacecraft level, Thales Alenia Space (Turin) responsible for the 
service module, and Astrium (Toulouse) responsible for the telescope, with in excess of a hundred subcontractors. PACS has been developed by a consortium of institutes led by MPE (Germany) and including 
UVIE (Austria); KU Leuven, CSL, IMEC (Belgium); CEA, LAM (France); MPIA (Germany); INAF-IFSI/OAA/OAP/OAT, LENS, SISSA (Italy); IAC (Spain). This development has been supported by the funding 
agencies BMVIT (Austria), ESA-PRODEX (Belgium), CEA/CNES (France), DLR (Germany), ASI/INAF (Italy), and CICYT/MCYT (Spain). SPIRE has been developed by a consortium of institutes led by Cardiff 
University (UK) and including Univ. Lethbridge (Canada); NAOC (China); CEA, LAM (France); IFSI, Univ. Padua (Italy); IAC (Spain); Stockholm Observatory (Sweden); Imperial College London, RAL, UCL-
MSSL, UKATC, Univ. Sussex (UK); and Caltech, JPL, NHSC, Univ. Colorado (USA). This development has been supported by national funding agencies: CSA (Canada); NAOC (China); CEA, CNES, CNRS 
(France); ASI (Italy); MCINN (Spain); SNSB (Sweden); STFC, UKSA (UK); and NASA (USA). HIPE is a joint development (are joint developments) by the Herschel Science Ground Segment Consortium, 
consisting of ESA, the NASA Herschel Science Center, and the HIFI, PACS and SPIRE consortia. Funding for SDSS-III has been provided by the Alfred P. Sloan Foundation, the Participating Institutions, the 
National Science Foundation, and the U.S. Department of Energy Office of Science. The SDSS-III web site is http://www.sdss3.org/. SDSS-III is managed by the Astrophysical Research Consortium for the 
Participating Institutions of the SDSS-III Collaboration including the University of Arizona, the Brazilian Participation Group, Brookhaven National Laboratory, Carnegie Mellon University, University of Florida, the 
French Participation Group, the German Participation Group, Harvard University, the Instituto de Astrofisica de Canarias, the Michigan State/Notre Dame/JINA Participation Group, Johns Hopkins University, 
Lawrence Berkeley National Laboratory, Max Planck Institute for Astrophysics, Max Planck Institute for Extraterrestrial Physics, New Mexico State University, New York University, Ohio State University, 
Pennsylvania State University, University of Portsmouth, Princeton University, the Spanish Participation Group, University of Tokyo, University of Utah, Vanderbilt University, University of Virginia, University of 
Washington, and Yale University. This publication makes use of data products from the Two Micron All Sky Survey, which is a joint project of the University of Massachusetts and the Infrared Processing and 
Analysis Center/California Institute of Technology, funded by the National Aeronautics and Space Administration and the National Science Foundation.

\bibliographystyle{mn2e}
\bibliography{references}

\end{document}